\documentclass[prb,twocolumn,showpacs,superscriptaddress]{revtex4}
\usepackage[latin1]{inputenc}
\usepackage{graphicx}
\usepackage{amssymb}
\usepackage{amsmath}
\usepackage{xspace}
\usepackage{dcolumn}
\usepackage{bm}

\newfont{\tensy}{cmsy10}
\newcommand{\chem}[1]{{$\fontdimen16\tensy=3.0pt
    \fontdimen17\tensy=3.0pt \mathrm{#1}$}}

\newcommand{\etal}[0]{{\it et al.}\@\xspace}
\newcommand{\ie}[0]{i.e.\@\xspace}
\newcommand{\eg}[0]{e.g.\@\xspace}
\newcommand{\rmi}{\mathrm{i}}

\newcommand{\las}[0]{\langle}
\newcommand{\ras}[0]{\rangle}

\begin{document}
{\scriptsize \em 6th.December 2006}

\title[Bipolaron dispersion in one dimension]%
{Polaron and bipolaron dispersion curves in one dimension for intermediate coupling}

\author{D. M. Eagles} 
\altaffiliation{%
Corresponding author.  Present address:
19, Holt Road, Harold Hill, Romford, Essex RM3 8PN, England.
email: d.eagles@ic.ac.uk}
\affiliation{%
  Instituto de Investigaciones en Materiales, Universidad Nacional
  Aut\'{o}noma de M\'{e}xico, Apartado Postal 70-360, 04510 M\'{e}xico D.F., 
  M\'{E}XICO}
\affiliation{%
  Department of Physics, University of Pretoria, Pretoria,
  0002 South Africa}

\author{R. M. Quick}
\affiliation{%
  Department of Physics, University of Pretoria, Pretoria,
  0002 South Africa}
\author{B. Schauer}
\affiliation{%
  Department of Physics, University of Pretoria, Pretoria,
  0002 South Africa}

\begin{abstract}

Bipolaron energies are calculated as a function of wave vector by a
variational method of Gurari appropriate for weak or intermediate
coupling strengths, for a model with electron-phonon interactions
independent of phonon wave vectors and a short-ranged Coulomb
repulsion.  It is assumed that the bare electrons have a constant
effective mass.  A two-parameter trial function is taken for the
relative motion of the two electrons in the bipolaron.  Energies of
bipolarons are compared with those of two single polarons as a function
of wave vector for various parameter values.  Results for effective
masses at the zone center are also obtained.  Comparison is made with
data of other authors for bipolarons in the Hubbard-Holstein model,
which differs mainly from the present model in that it has a
tight-binding band structure for the bare electrons.

\end{abstract}


\maketitle



%
%
%
\section{\label{sec:introduction}Introduction}
%
%
%

Many authors have studied energies of
large\cite{Vi61,HiTo85,Ad89,VePeDe91,BaGeIaNi91,EmHi89,DzBaAb96,BaKhSh97,%
dCPe98,PoFoDeBaKl00,My01,Mu02,SeEr02,Ka04,KaLaSy03} and
small\cite{ChSc76,AlRa81,CoEcSo84,RaTh92,smallbipolarons,AlMo94,Al96,AlBr99,
AlKo02,Marsiglio95,
FeRoWeMi95,WeRoFe96,FeLoWe00,LaMaPu97,FiKu97,PrAu99,FrWa99,Sil99,ZhJeWh99,
BoKaTr00,BoTr01,
ElShBoKuTr03}\newline\cite{IaPeCadF01,deFiCaIaMaPeVe01,Ma04,HoAivdL04,HovdL05,Zh05}
bipolarons in various numbers of dimensions, and some have made
calculations of bipolaron effective masses at the band
minimum.\cite{EmHi89,dCPe98,PoFoDeBaKl00,CoEcSo84,LaMaPu97,Sil99,BoKaTr00,
BoTr01,
ElShBoKuTr03} However, we are not aware of published calculations of
bipolaron energies as a function of wave vector which extend to large
wave vectors except in the case of small bipolarons when the
electron-phonon coupling is strong.\cite{FeRoWeMi95,WeRoFe96,FeLoWe00}
In this paper we take up a study of this problem in a one-dimensional
model with local interactions by use of a variational method used by
Gurari for the single-polaron problem.\cite{Gu53}  The method is
appropriate for intermediate electron-phonon coupling strengths.  It
was discussed in some detail in a review article by
Fr\"{o}hlich,\cite{Fr54} and, for large polarons, gives the same
results for binding energies and effective masses as those obtained by
Lee \etal\cite{LePi52,LeFlPi53} using different approaches.  The
Hamiltonian we shall use for the bipolaron problem is similar but not
identical to that of the Hubbard-Holstein
model,\cite{Hu63,Ho59a,Al96,AlKo02} and will be formulated in terms of
center-of-mass and relative coordinates of two electrons rather than in
terms of electron creation and annihilation operators.   This permits
us to follow the variational method for the single-polaron problem with
only minor modifications.  The biggest difference of our model from the
Hubbard-Holstein model is our assumption of a constant bare mass for
the electron.  We have not yet found a way to apply the Gurari method
to the Hubbard-Holstein model itself.

The original motivation for this work was to help to find out whether large
enhancements of electron-electron attractions mediated via phonons or
other intermediate bosons predicted in the simplest perturbation
approach to interactions\cite{Pa59,Ea66,Ea94,Ea94_2,Ho65} may still occur
when complications due
to intermediate coupling are included.  For metals with Fermi energies
large compared with the energy of any boson mediating electron-electron
attractions, Eliashberg theory has been used to show that the net
effect of enhancements of attractions is unlikely to lead to
high-temperature superconductivity at high currents either in
three-dimensional\cite{Ho65} or quasi one-dimensional systems.\cite{Ea94_2}
However, for small Fermi energies, infinite enhancements are predicted by the
simplest perturbation  approach at some drift velocities in one
dimension.\cite{Ea94,Ea94_2} Study of the bipolaron problem in one dimension
should give some insight into how this conclusion is modified by
effects beyond perturbation theory in the limit of low concentrations
of electrons.  Although we shall use the term ``phonon'' for the boson
mediating the attraction, we have in mind that interactions mediated by
plasmons may be of practical importance for understanding possible
high-temperature superconductivity associated with bipolarons.

There have been many published studies of bipolarons in one dimension
(including studies of two-site models),\cite{RaTh92,PoFoDeBaKl00,CoEcSo84,Al96,
AlKo02,
Marsiglio95,FeRoWeMi95,WeRoFe96,FeLoWe00,FiKu97,FrWa99,Sil99,ZhJeWh99,
BoKaTr00,
BoTr01,ElShBoKuTr03,deFiCaIaMaPeVe01,HoAivdL04,HovdL05}
but these mostly concentrate on finding the energies for the ground state as a function
of coupling strength and Coulomb repulsion.  Some calculate the
bipolaron effective mass at the bottom of the band, but most do not
discuss how energies vary with center-of-mass wave vector well away
from the band minimum, except for strong electron-phonon coupling.
Hohenadler \etal\cite{HoAivdL04} give graphical results for the spectral
function of bipolarons as a function of wave vector for various cases
where the coupling is not very strong, and approximate $E(k)$ curves can
be deduced from these.  An easier comparison to make is with some
unpublished calculations performed by S. El Shawish for the Holstein
model for some parameters corresponding roughly to some of those we
have used.  These values correspond to high ratios of phonon energy to
electronic transfer integral $t$. 

In Sec.~\ref{sec:model} we introduce the Hamiltonian and our variational method.
Some numerical results are presented in Sec.~\ref{sec:results}, and some
discussion is given in Sec.~\ref{sec:discussion}.

%
%
%
\section{\label{sec:model}Hamiltonian and variational method}
%
%
%

With a notation somewhat similar to that of Ref.~\onlinecite{BaGeIaNi91} but
modified to apply to one dimension and for short-range interactions, we write the
Hamiltonian $H$ for the bipolaron problem with constant bare effective
masses in the form
\begin{equation}\label{eq:model}
H = H_X + H_u +H_p +H_\text{e-p}.
\end{equation}
Here $H_X$ is the center-of-mass kinetic energy given by
\begin{equation}
H_X= -\mbox{\small$\frac{1}{2}$}\nabla_X^2,
\end{equation}
where the center-of-mass coordinate
\begin{equation}
X = \mbox{\small$\frac{1}{2}$}(x_1+x_2),
\end{equation}
and $x_1$, $x_2$ are the coordinates of the two electrons;
$H_u$ is the Hamiltonian for relative motion, with
\begin{equation}\label{eq:Hu}
H_u = -2\nabla_u^2 + W(u),
\end{equation}
where
\begin{equation}
u = x_1-x_2
\end{equation}
is the relative coordinate and
\begin{equation}
W(u) = 
\begin{cases}\label{eq:P}
  P &-\mbox{\small$\frac{1}{2}$}a < u <\mbox{\small$\frac{1}{2}$}a\,,\\
  0  &\text{otherwise}\,,
\end{cases}
\end{equation}
with $a$ the lattice constant; the phonon Hamiltonian $H_\text{p}$ and the
electron-phonon interaction $H_\text{e-p}$ are given by
\begin{equation}
H_\text{p} = \sum_k a_k^{\dagger} a^{\phantom{\dagger}}_k
\end{equation}
and
\begin{equation}\label{eq:H_e-p}
H_\text{e-p} = -iV(a/L)^{\frac{1}{2}} \sum_k [2 \cos(\mbox{\small$\frac{1}{2}$}ku)
e^{\rmi kX}a_k] + \text{H.c.},
\end{equation}
where $a_k^{\dagger}$ and $a^{\phantom{\dagger}}_k$ are creation and destruction operators
for phonons of wave number $k$  and $L$ is the length of the crystal.
We use the usual reduced units, with units of energy, length and mass
equal to $\hbar \omega$, $(\hbar/2 m \omega)^{\frac{1}{2}}$ and $2m$
respectively, where $\omega$ is the phonon angular frequency and $m$
is the bare electron mass.  The form of the potential term in $H_u$ is
similar to but not the same as that in the Hubbard model because, with
the form used, two electrons within a unit cell do not always interact,
but this is compensated by interactions between electrons which are in
neighbouring cells but separated by less than $a$. We do not include any
spin-dependent terms in the Hamiltonian, and in the following we shall
not include any terms involving electron-spin wave functions.
For bipolarons we implicitly assume that the two electrons in
the pair have opposite spin by choosing a wave function for relative
motion which is even in the relative coordinates.

For a given center-of-mass vector $Q$, we adopt a trial wave function $\Psi$ 
of the form
\begin{equation}
\Psi = L^{-\frac{1}{2}} e^{\rmi QX} \phi_Q(u) \prod_k \psi(Q,k,u) \chi,
\end{equation}
where $\chi$ is the phonon or other boson vacuum,
\begin{eqnarray}
\psi(Q,k,u) &=& N(Q,k,u)\\
&&\times\left[ 1 + L^{-\frac{1}{2}} c(Q,k) \cos(\mbox{\small$\frac{1}{2}$}
ku) e^{-\rmi kX} a_k^\dagger \right],\nonumber
\end{eqnarray}
where $c(Q,k)$ are variational parameters, $\phi_Q(u)$ is a normalised 
even function, 
\begin{equation}
\int |\phi_Q(u)|^2 du = 1,
\end{equation}
depending on one or more parameters, and 
$N(Q,k,u)$ is a normalisation constant given by
\begin{equation}
N(Q,k,u) = \left[ 1 + L^{-1} |c(Q,k)|^2 \cos^2 (\mbox{\small$\frac{1}{2}$} ku) \right]^{-
\frac{1}{2}} \simeq 1.
\end{equation}

With such a trial wave function, calculations similar to those in Sec.~4 of 
Fr\"{o}hlich's review article\cite{Fr54} for single polarons give the 
expectation values of the different terms in $H$.  After replacing sums by
integrals with use of the replacement
\begin{equation}
\sum_k \mapsto \frac{L}{2\pi} \int_{-\pi/a}^{\pi/a} dk,
\end{equation}
we find
\begin{equation}\label{eq:<H_p>}
\langle H_\text{p} \rangle = \frac{1}{2\pi}\int_{-\pi/a}^{\pi/a} |c(Q,k)|^2 h_k dk,
\end{equation}
where
\begin{equation}\label{eq:h_k}
h_k = \int |\phi_Q(u)|^2 \cos^2 (\mbox{\small$\frac{1}{2}$}ku) \: du,
\end{equation}
\begin{equation}\label{eq:<H_p>2}
\langle H_\text{e-p} \rangle = \frac{i}{\pi} V a^\frac{1}{2} \int_{-\pi/a}^{\pi/a} 
[c^*(Q,k)-c(Q,k)] h_k dk,
\end{equation}
\begin{eqnarray}\label{eq:<H_X>}
    \langle H_X \rangle &=& \frac{1}{2} Q^2 
    - \frac{1}{2\pi} \int_{-\pi/a}^{\pi/a} Qk  |c(Q,k)|^2 h_k \\\nonumber
    &+& \frac{1}{4\pi} \int_{-\pi/a}^{\pi/a} |c(Q,k)|^2 k^2 h_k dk\\\nonumber
    &+& \frac{1}{8\pi^2} 
    {\int\int}_{-\pi/a}^{\pi/a} kk^\prime |c(Q,k)|^2 
    |c(Q,k^\prime)|^2 h_{kk^\prime} dk dk^\prime
    \,.
\end{eqnarray}
Here
\begin{equation}\label{eq:hkk}
h_{kk^\prime} = \int |\phi_Q(u)|^2 \cos^2 (\mbox{\small$\frac{1}{2}$}ku) 
\cos^2 (\mbox{\small$\frac{1}{2}$} k^\prime u) \: du
\end{equation}
and
\begin{equation}\label{eq:<H_u>}
\las H_u \ras = E_u +  \frac{1}{4\pi} \int_{-\pi/a}^{\pi/a} |c(Q,k)|^2 k^2 (1-h_k) dk,
\end{equation}
where the first term is independent of the variational coefficients $c(Q,k)$.
Going back to sums over k in
Eqs.~(\ref{eq:<H_p>}),~(\ref{eq:<H_p>2}),~(\ref{eq:<H_X>})
and~(\ref{eq:<H_u>}), by minimisation of $\las H\ras$ with respect to
$c(Q,k)$ and $c^*(Q,k)$, we find
\begin{equation}
c(Q,k) = \frac{-2iV a^\frac{1}{2}}{1 - (Q  - h_k^{-1} G_k)k 
+ \mbox{\small$\frac{1}{2}$}h_k^{-1} k^2},
\end{equation}
where
\begin{eqnarray}
  G_k 
  & = &
  \frac{1}{2\pi} \int_{-\pi/a}^{\pi/a} k^\prime h_{kk^\prime} 
  |c(Q,k^\prime)|^2 dk^\prime \\
  & = &\nonumber
  \frac{2V^2 a}{\pi}  \int_{-\pi/a}^{\pi/a} \frac{k^\prime h_{kk^\prime}}
  {\left[1-(Q  - h_{k^\prime}^{-1} G_{k^\prime})k^\prime + \mbox{\small$\frac{1}{2}$}
  h_{k^\prime}^{-1} k^{\prime \:2}\right]^2} dk^\prime.
\end{eqnarray}
From Eqs.~(\ref{eq:h_k}) and~(\ref{eq:hkk}),
\begin{equation}\label{eq:h_kk2}
h_{kk'}=\mbox{\small$\frac{1}{2}$}h_k+\mbox{\small$\frac{1}{2}$}h_{k'}+\mbox{\small$\frac{1}{4}$}
h_{k+k'}+\mbox{\small$\frac{1}{4}$}h_{k-k'}-\mbox{\small$\frac{1}{2}$}.
\end{equation}
For the purpose of determining the coefficients we will make the following 
approximation
\begin{equation}\label{eq:h_kk3}
h_{kk'} \simeq h_k h_{k'}.
\end{equation}
This gives
\begin{equation}\label{eq:c(Q,k)}
c(Q,k) 
\simeq 
\frac{-2iV a^\frac{1}{2}}{1 - (Q  - g)k + \mbox{\small$\frac{1}{2}$} h_k^{-1} k^2},
\end{equation}
where
\begin{equation}\label{eq:g}
  g= 
  \frac{2V^2 a}{\pi}  \int_{-\pi/a}^{\pi/a} \frac{k^\prime h_{k^\prime}}{\left[1-
  S k^\prime
  + \mbox{\small$\frac{1}{2}$} h_{k^\prime}^{-1} k^{\prime \:2}\right]^2} dk^\prime,                                 
\end{equation}
with
\begin{equation}
S = Q - g.
\end{equation}
From Eqs.~(\ref{eq:model}),~(\ref{eq:<H_p>}),~(\ref{eq:<H_p>2}),~(\ref{eq:<H_X>}),~(\ref{eq:<H_u>})
and~(\ref{eq:c(Q,k)}), we find that the expectation value of
$H$, which we write as $E(Q)$, is given by
\begin{widetext}
\begin{eqnarray}\label{eq:E(Q)}\nonumber
E(Q) 
&=& \mbox{\small$\frac{1}{2}$} (Q^2-g^2) 
     -\frac{2V^2 a}{\pi} 
     \int_{-\pi/a}^{\pi/a}
     \frac{h_k}{1 - Sk + \frac{1}{2} h_{k}^{-1} k^2} dk
\\  
&&+  
     \frac{2V^4 a^2 }{\pi^2} 
     \int_{-\pi/a}^{\pi/a} \int_{-\pi/a}^{\pi/a} 
     \frac{kk^\prime (h_{kk^\prime}-h_k h_{k\prime})}
     {(1- Sk + \frac{1}{2} h_{k}^{-1} k^2)^2
      (1-Sk^\prime + \frac{1}{2}h_{k^\prime}^{-1} k^{'\:2})^2}
      dk dk^\prime + E_u\,.
\end{eqnarray}
\end{widetext}
Although the variational coefficients determined in this approximation 
do not represent the optimal choice, nonetheless the expectation value of 
$H$, $E(Q)$, yields an upper bound to the exact energy.

The part $E_u$ of the expectation value of $H_u$ in Eq.~(\ref{eq:<H_u>}) which is
independent of $c(Q,k)$ can be written as
\begin{equation}\label{eq:E_u}
E_u=T_u+V_u.
\end{equation}
Here the relative kinetic energy $T_u$ is given by
\begin{equation}\label{eq:T_u}
T_u = \frac{2}{2\pi}\int |f_k|^2 k^2 dk,
\end{equation}
where $f_k$ is given by
\begin{equation}\label{eq:f_k}
f_k= \int\phi(u) e^{\rmi ku}du.
\end{equation}
In Eq.~(\ref{eq:T_u}) the integral over $k$ is, in principle, unrestricted, 
\ie is an integral over $k$ from $-\infty$ to $\infty$. 
However, we shall restrict our trial wave
functions for relative electron motion to those that do not have
Fourier components for $k>k_m=\pi/a$, and for such trial wave functions
we use an integral from $-k_m$ to $k_m$.

Writing
\begin{equation}\label{eq:d_k}
\int_{-\infty}^{\infty}|\phi(u)|^2 e^{\rmi ku}du=d_k
\end{equation}
we have, for a trial function for which $f_k=0$ for $|k|>k_m$,

\begin{equation}\label{eq:d_k2}
  d_k
  =
  \frac{1}{2\pi}
  \int_{-(k_m-\frac{1}{2}|k|)}^{k_m-\frac{1}{2}|k|}
  f^{*}_{k'-\frac{1}{2}k}f_{k'+\frac{1}{2}k}dk'.
\end{equation}

Using
\begin{equation}
\int_{-\frac{1}{2}a}^{\frac{1}{2}a} e^{-\rmi ku}du 
=
\frac{2}{|k|}{\rm sin}(\mbox{\small$\frac{1}{2}$}|k|a),
\end{equation} 
from Eq.~(\ref{eq:P}) we find that the potential term $V_u$ in Eq.~(\ref{eq:E_u}) is given by
\begin{equation}\label{eq:V_u}
V_u
=
\frac{2P}{2\pi}
\int_{-2k_m}^{2k_m} d_k[{\rm sin}
(\mbox{\small$\frac{1}{2}$}ka)/k]dk
\end{equation}
We restrict the integral to the limits shown in Eq.~(\ref{eq:V_u})
because $d_k$ vanishes for $|k|>2k_m$.
Remembering that 
\begin{equation}
\cos^2(\mbox{\small$\frac{1}{2}$}ku) = \mbox{\small$\frac{1}{2}$} [\cos(ku)+1],
\end{equation}
Eqs.~(\ref{eq:h_k}) and~(\ref{eq:d_k}) enable us to write

\begin{equation}\label{eq:h_k2}
h_k=\mbox{\small$\frac{1}{2}$}+\mbox{\small$\frac{1}{2}$}d_k;
\end{equation}
$h_{kk'}$ is then determined from Eq.~(\ref{eq:h_kk2}).

We now consider a two-parameter trial function for the 
relative motion, with an assumed function in real space modified by
replacement of all Fourier components with wave vectors of magnitude
greater than $k_m$ replaced by zero. 

\subsection*{Pseudo real-space function for relative motion}

We consider a function which has Fourier transforms 
up to $|k|=k_m$ of the same form, up to a proportionality
factor, as  the transforms of $\phi$, where
\begin{equation}\label{eq:phi(u)}
\phi(u) = N_0(1+b|u|)e^{-\lambda|u|},
\end{equation}
with $b$ and $\lambda$ adjustable parameters, and $N_0$ a normalisation
factor.   
If $b$ in the trial function is small, then the maximum
of $\phi$ is at $u=0$, whereas if
$b>\lambda$ there are two maxima at finite $|u|$. 
For $|k|>k_m$, we
assume that the Fourier transform of $\phi$ is zero.
Thus, using Eq.~(\ref{eq:phi(u)}), $f_k$ of Eq.~(\ref{eq:f_k}) is given by
\begin{equation}
f_k =
\begin{cases} 
  N\left[
    \frac{2\lambda}{(\lambda^2+k^2)}
    +\frac{2b(\lambda^2-k^2)}{(\lambda^2+k^2)^2}
    \right]
    &\text{if $|k|<k_m$\,,}\\
    0 &\text{if $|k|>k_m$}\,,
\end{cases}
\end{equation}
where $N$ is a normalisation
factor.  From $(1/2\pi)\int_{-k_m}^{k_m} |f_k|^2=1$, we find $N$ is given by
\begin{equation}\label{eq:N^2}
N^2 =B^{-1} N_0^2,
\end{equation}
where
\begin{equation}\label{eq:N_0}
N_0=(1/\lambda + b/\lambda^2 +b^2/2\lambda^3)^{-\frac{1}{2}}
\end{equation}
is the normalisation factor of a wave function given by Eq.~(\ref{eq:phi(u)})
before imposing a restriction on the Fourier components for $|k|>k_m$,
and
\begin{equation}\label{eq:B}
  B
  =
  \frac{1}{2\pi}\int_{-k_m}^{k_m}|f_{0k}|^2,
\end{equation} 
where 
$f_{0k}$ is the same as $f_k$ but with the normalisation factor replaced
by $N_0$.
The departure from unity of the integral of Eq.~(\ref{eq:B}) gives the
fractional change of the square of the normalisation factor due to
putting the Fourier transforms of $\phi$ for $|k|>k_m$ as zero.

From Eqs.~(\ref{eq:h_k}) and~(\ref{eq:phi(u)}), if one were to suppose that the assumption of
zero Fourier transforms of $\phi$ for $|k|>k_m$ only affected $h_k$ for
$|k|<k_m$ via the change of normalisation factor, we would find
\begin{equation}\label{eq:h_k/N^2}
  \frac{h_k}{N^2} 
  = 
  \frac{2\lambda}{4\lambda^2+k^2}+\frac{2b(4\lambda^2-k^2)}{(4\lambda^2+k^2)^2}
  +\frac{2b^2(8\lambda^3-6\lambda k^2)}{(4\lambda^2+k^2)^3}+\mbox{\small$\frac{1}{2}$}
\end{equation}
for $|k|\leq k_m$. 
We have verified that, if the integral in Eq.~(\ref{eq:d_k2}) is extended to be from
$-\infty$ to $\infty$ then Eq.~(\ref{eq:h_k2}) gives the result of Eq.~(\ref{eq:h_k/N^2}).

The kinetic term $T_u$ in Eq.~(\ref{eq:E_u}) is given by Eq.~(\ref{eq:T_u}).
If $\phi$ were taken as in Eq.~(\ref{eq:phi(u)}) without removal of the
Fourier components for $|k|>k_m$, then using Eqs.~(\ref{eq:Hu}),~(\ref{eq:P})
and~(\ref{eq:phi(u)}), the potential term $V_u$ in Eq.~(\ref{eq:E_u}) would be
\begin{eqnarray}\label{eq:V_u2}\nonumber
  V_u
  &=&
  P\int_{-\frac{1}{2}a}^{\frac{1}{2}a}{|\phi|^2 du}\\\nonumber
  &=&
  P N_0^2[(1/\lambda+b/\lambda^2+b^2/2\lambda^3)(1-e^{-\lambda a})\\
  &&-(ab/\lambda+b^2 a/2\lambda^2-b^2 a^2/4\lambda)e^{-\lambda a}]\,.
\end{eqnarray}
By comparison of numerical results obtained from Eq.~(\ref{eq:V_u}) and~(\ref{eq:h_k2}) 
with those from Eqs.~(\ref{eq:h_k/N^2}) and~(\ref{eq:V_u2}), we can find
the effect of putting Fourier components of $\phi$ for $|k|>k_m$ on
$h_k$ and $V_u$ for given values of the parameters.

%
%
%
\section{\label{sec:results}Numerical results}
%
%
%

Marsiglio\cite{Marsiglio95} states that the physical region of the Hubbard-Holstein
model requires a condition $\alpha^2/K < U$, where $\alpha$ is a factor
multiplying local vibrational coordinates $x_i$ in the electron-phonon
interaction, and $K$ appears in the expression $\frac{1}{2}Kx_i^2$ for
the vibrational potential energy.  If we identify $Pa$ in our model with
$U$, although this correspondence is not exact, the same type of
condition would require that
\begin{equation}\label{eq:V-Pa}
2 V^2 \leq Pa.
\end{equation}
However, in attractive Hubbard models ($U<0$), the negative $U$ is usually
thought to be brought about by effects of electron-phonon interactions
overcoming a positive Hubbard $U$, and Marsiglio's condition would
appear to imply that negative $U$ Hubbard models are unphysical.
Therefore we shall not restrict our numerical work to the region
defined by Eq.~(\ref{eq:V-Pa}), although in cases in which we are using our model
with short-range forces as a simple approximation for effects of
long-range forces we shall need to take the condition more seriously.

For our numerical calculations we consider two types of cases.  First
we choose parameters which may be applicable to oxidised atactic
polypropylene (OAPP),\cite{fn0,fn} and make use of some of the
parameters used previously in the model for superconductivity in
channels composed of arrays of quasi one-dimensional
filaments.\cite{Ea94_2}  As in Ref.~\onlinecite{Ea94_2}, we take the
cross section of individual filaments to be 0.25 nm$^2$. However,
because the periodic potential due to aligned dipoles near the strings
of charges forming the filaments according to the model of Grigorov
\etal\cite{Gr90,GrAnSm90,Gr91} for individual filaments may not have
very deep minima, we assume here that the bare electron mass is
$m_\text{e}$ and not $2m_\text{e}$ as in Ref.~\onlinecite{Ea94_2}.  We
consider two variants of parameters which could be applicable to OAPP.
One, as in Ref.~\onlinecite{Ea94_2}, where the excitations mediating
the electron-electron attraction are plasmons, and the other where
optical phonons of average energy 0.36 eV mediate the attraction.  For
the plasmon-induced interaction, we should strictly have long-ranged
forces, but we hope that the model used here with short-range forces
will give a first approximation to the real situation.  We choose 0.36
eV for a phonon energy because there are several branches of the phonon
spectrum associated with \chem{C-H_2} and \chem{C-H_3} stretching
vibrations whose energies at long wavelengths lie between 0.35 and 0.37
eV.\cite{McWa61}

For given values of $V^2$, $a$, $Q$ and $P$, we solve for $g$ of
Eq.~(\ref{eq:g}) and then minimise the total energy with respect to
$\lambda$ and $b$, using our full expressions of Eqs.~(\ref{eq:d_k2})
and~(\ref{eq:h_k2}) for $h_k$, Eq.~(\ref{eq:V_u}) for $V_u$, and the
normalisation factor of Eq.~(\ref{eq:N^2}) for $f_k$.  For
computational purposes, we made use of programs or modifications of
programs from a book.\cite{numrec}

In the model for strings or nanofilaments of
Refs.~\onlinecite{Gr90,GrAnSm90,Gr91}, if there is a
periodic potential acting on the electrons in the string, it is due to
groups of about three aligned dipoles surrounding each electron, and so
the period of the potential, or lattice constant, $a$, is equal to the
inverse of the linear electron concentration $c = nd^2$, where $n$ is
the three-dimensional concentration in the filament and $d^2$ is its
cross section.  Thus we may not be free to choose the carrier
concentration and lattice constant independently. However, we note that
a recent theoretical study of channels through films of oxidised atactic
polypropylene making use of Bose condensation of bosons in an array
of nanofilaments with an $E(K)$ curve for bosons consisting of a
combination of linear and quadratic terms,\cite{Ea05} as indicated to
occur in studies of Cooper pair dispersion,\cite{AdCaPuRiFoSoDLVaRo00}
did not assume a periodic potential acting on the electrons in the
nanofilaments.  Another constraint was imposed in
Ref.~\onlinecite{Ea05} because it was thought that it was probably
necessary for the Fermi energy to be smaller than a quarter of the
energy of the excitation mediating the electron-electron attraction in
order to have the possibility of large enhancements of interactions at
high drift velocities.\cite{Ea94_2}  However, in view of the results of
the present paper, these enhancements may not occur when calculations
of electron-electron interactions go beyond second-order perturbation
theory.

Bearing in mind the possible constraints, and assuming $d^2=0.25$
nm$^2$ and $m_\text{b} = m_\text{e}$ as discussed above, we find two
values of $a$ and the related carrier concentrations corresponding to
our two possible choices of phonons or plasmons to mediate the
attraction.   In both cases we use parameters such that the ratio of
$a$ to the polaron radius $r_\text{p} = (\hbar/2m\omega)^{\frac
{1}{2}}$ satisfies $a/r_\text{p}=4$.  Using a value for the
high-frequency dielectric constant of 2.2,\cite{dielectric} we find,
for plasmons mediating the attraction, that $a/r_\text{p} = 4$ implies
$a= 0.53$ nm, the linear concentration $c=1/a=1.9\times 10^7$
cm$^{-1}$, the three-dimensional carrier concentration $n$ within a
filament is $n= 7.6 \times 10^{21}$ cm$^{-3}$, the plasmon energy
$\hbar \omega_\text{p}$ calculated using a three-dimensional formula
appropriate for not too long wavelengths is $\hbar \omega_\text{p} =$
2.2 eV, the polaron radius $r_\text{p} = 0.132$ nm, and, from a
one-dimensional formula, the bare Fermi energy $\epsilon_F = 0.34$
eV.   The values of most of these quantities are only slightly
different from those used in Ref~\onlinecite{Ea94_2}.  For phonons of
energy $0.36$ eV, $a/r_\text{p} = 4$ implies $a = 1.3$ nm, $r_\text{p}
= 0.325$ nm, $n = 3.1 \times 10^{21}$ cm$^{-3}$ and $\epsilon_F=0.056$
eV.

We also consider a different type of case appropriate for a quantum
wire of a crystalline material.   If, \eg, we assume that $m=2m_\text{e}$
and $\hbar \omega$ = 0.05 eV, then the polaron radius is 0.62 nm, and a
ratio $a/r_\text{p} = 0.5$ would then imply a plausible value of the
lattice constant of 0.31 nm.   We note that, for such a small value of
$a/r_\text{p}$, there is no point in doing calculations for very large values
of $Q$, since the type of variational method we are using will not be
appropriate when the single-polaron energy lies more than $\hbar
\omega$ above the bottom of the band.\cite{Fr54}

Note that, when the polaron or bipolaron energy above the bottom of the
band becomes close to $\hbar \omega$, the discrepancy in energy from
the types of states more generally discussed (see
\eg Refs.~\onlinecite{FeRoWeMi95,WeRoFe96,FeLoWe00,ElShBoKuTr03}) becomes
large.  This is because our method requires one or two electrons (for
polarons or bipolarons) whose average wave vector or
centre-of-mass wave vector is equal to the centre-of-mass wave vector
of the polaron or bipolaron to be present whatever the energy of the
state, whereas more commonly used methods find the lowest energy of the
electron-phonon system for a given centre-of-mass wave vector.  Such
wave functions have only a small electron component at the wave vector
concerned when the threshold energy for emission of phonons is
approached.  We think that our method is concerned with states of more
physical interest than the states usually discussed in this energy
region.  These states, with the wave vector provided by phonons, and
electrons at the bottom of the band, do not help in describing what
happens when a polaron or bipolaron is accelerated rapidly past the
threshold for emission of phonons.

\begin{figure}
  \begin{center}
    \includegraphics[width=0.45\textwidth]{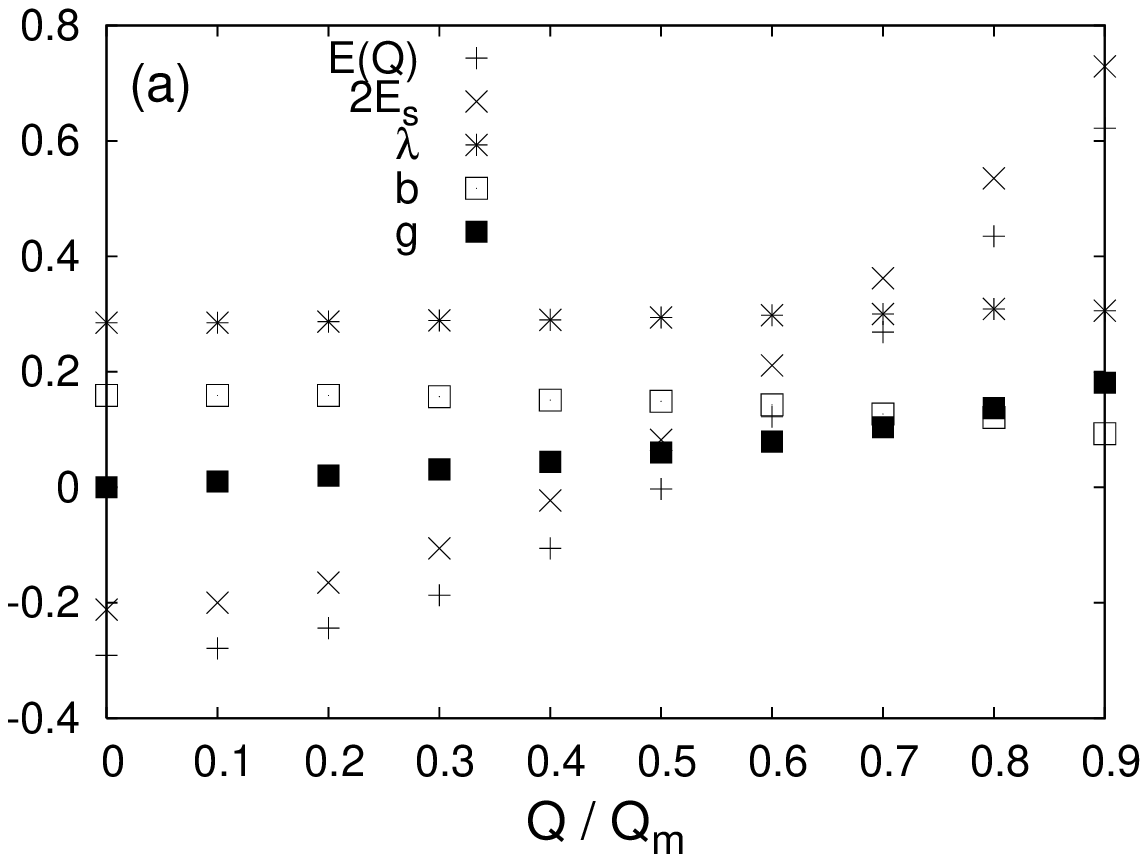}
    \includegraphics[width=0.45\textwidth]{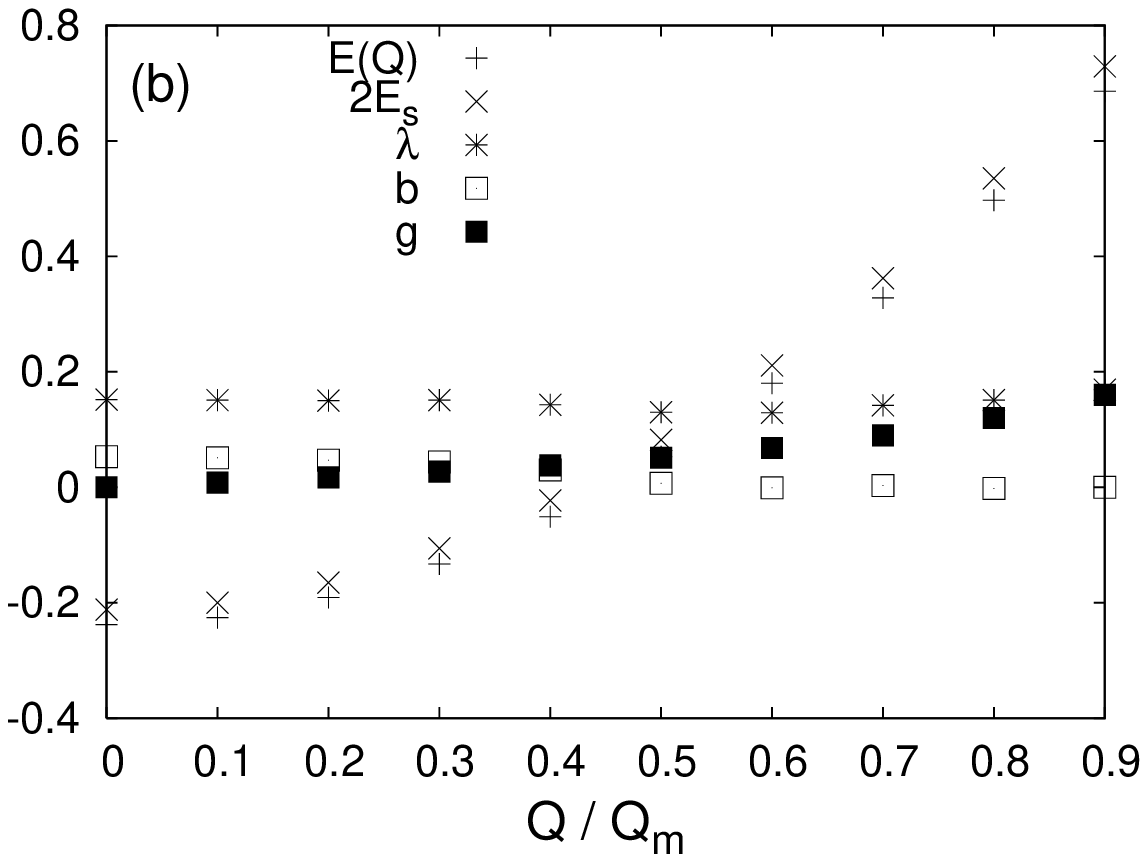}
    \caption{\label{fig:fig1}
      Bipolaron energies $E(Q)$, the energy $2E_\text{s}$ of two widely
      separated single polarons, the parameters $\lambda$ and $b$ of the
      trial wave function for relative motion of Eq.~(\ref{eq:phi(u)}), and the  quantity
      $g$ of Eq.~(\ref{eq:g}) as a function of bipolaron wave vector $Q$, for
      $V^2=0.125$, $a=4$, and (a) $P=0$, (b) $P=0.125$.  Energies are
      measured with respect to the bottom of the bare band, in units of the
      phonon energy. $Q_m$ is defined by Eq.~(\ref{eq:Q_m}).}
  \end{center}
\end{figure}

\begin{figure}
  \begin{center}
    \includegraphics[width=0.45\textwidth]{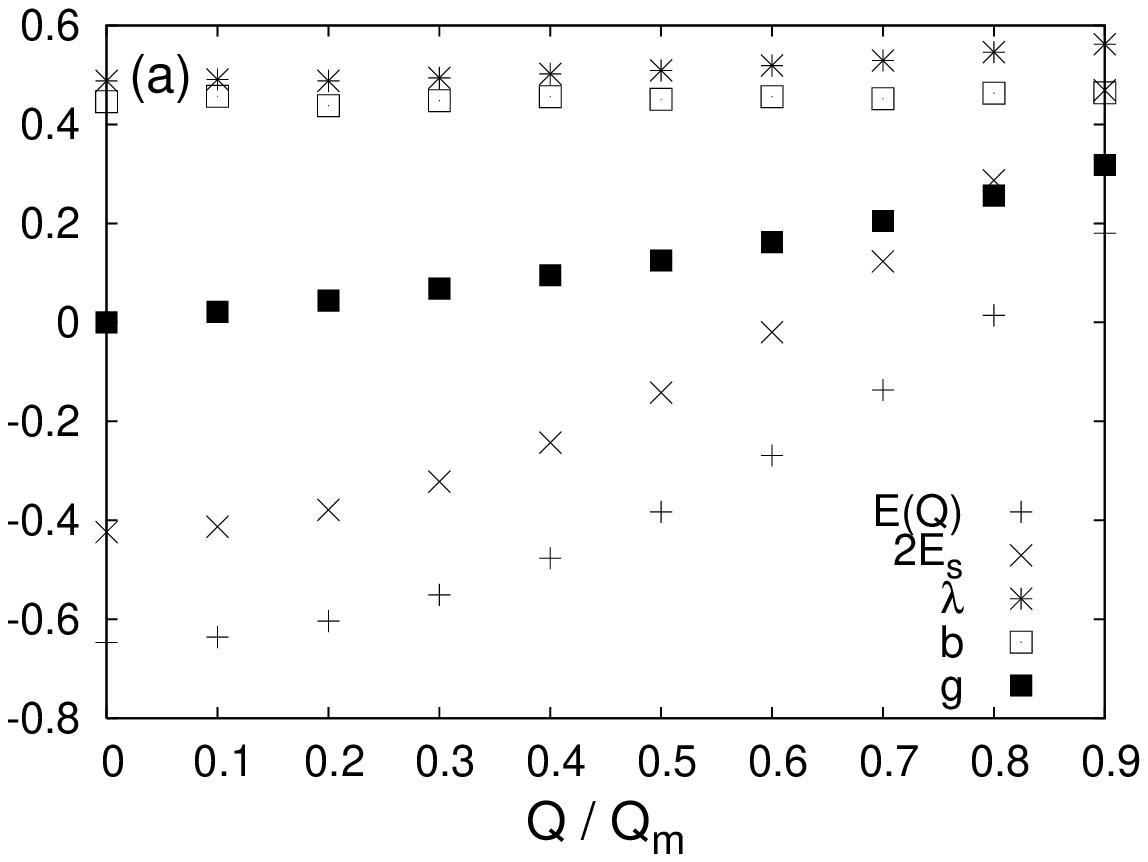}
    \includegraphics[width=0.45\textwidth]{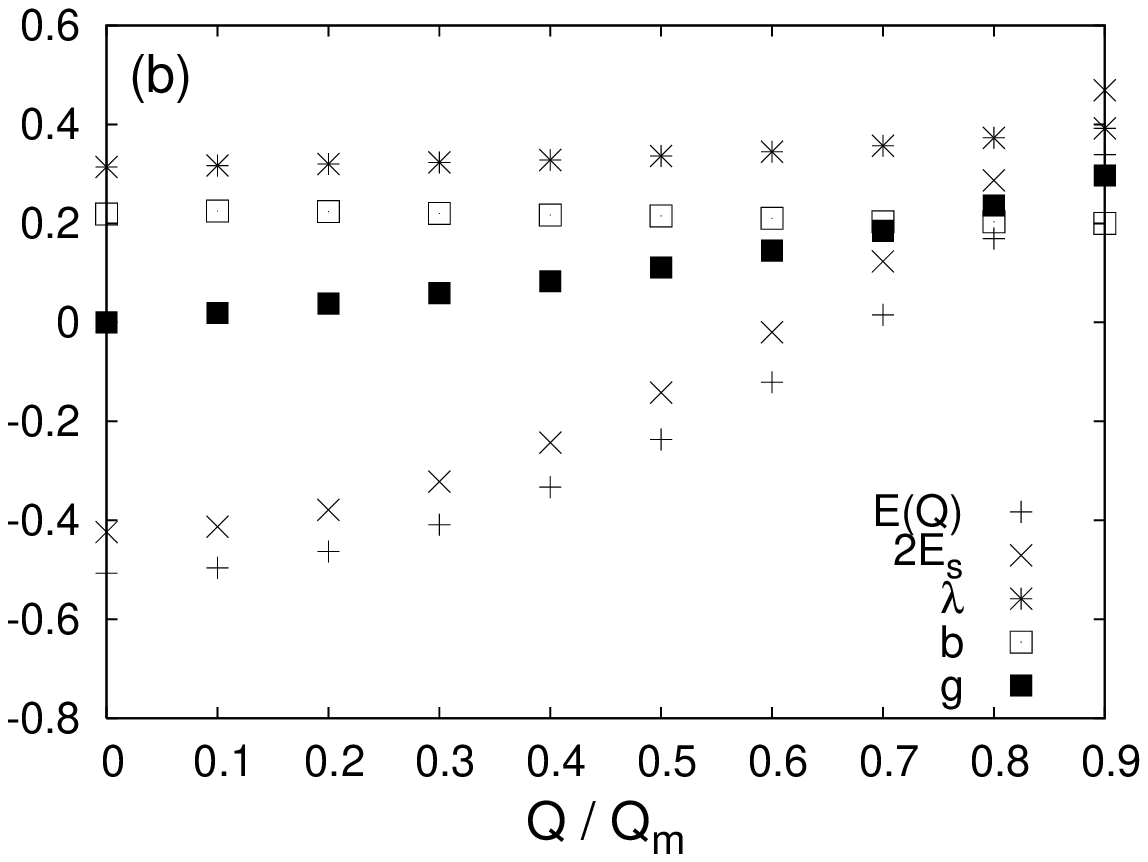}
    \caption{\label{fig:fig3}
      As for Fig.~\ref{fig:fig1}, but with $V^2=0.25$, $a=4$ and (a) $P=0$, (b)
      $P=0.25$.
    }
  \end{center}
\end{figure}

Since we are using $a/r_\text{p} =4$ for both possible parameter choices for
OAPP, the same computer calculations can be used for both plasmon or
phonon-induced interactions, with only the values of quantities obtained
in real units being different.  Figures~\ref{fig:fig1} and~~\ref{fig:fig3}
show values of the bipolaron energy $E(Q)$, the energy $2E_s$ of two widely
separated polarons each with wave vector $0.5Q$, and the
parameters $\lambda$ and
$b$ in the bipolaron trial function, 
for two values of $V^2$ and two related values of $P$ for each $V$.  
as a function of $Q/Q_m$, where 
\begin{equation}\label{eq:Q_m}
Q_m=2k_m=2\pi/a.
\end{equation}
For small $b$, $1/\lambda$ is the bipolaron radius in units of the polaron
radius.

\begin{figure}
  \begin{center}
    \includegraphics[width=0.45\textwidth]{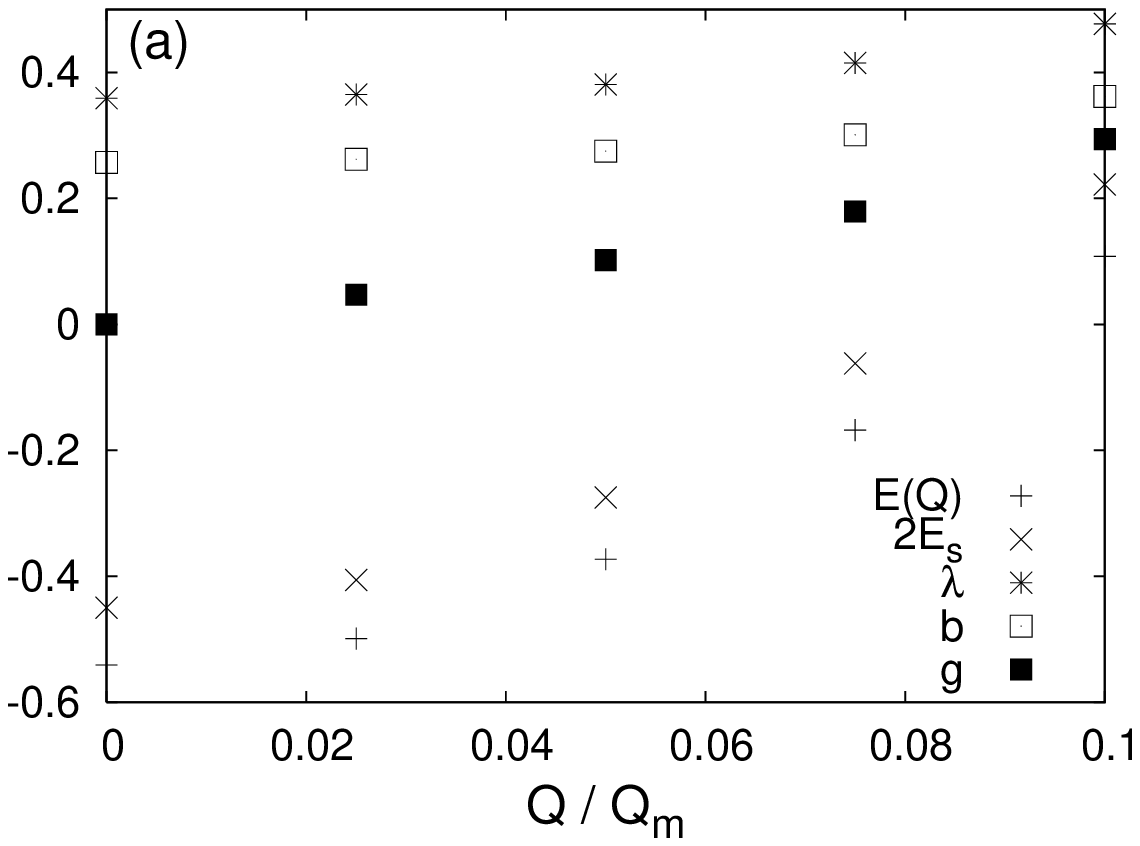}
    \includegraphics[width=0.45\textwidth]{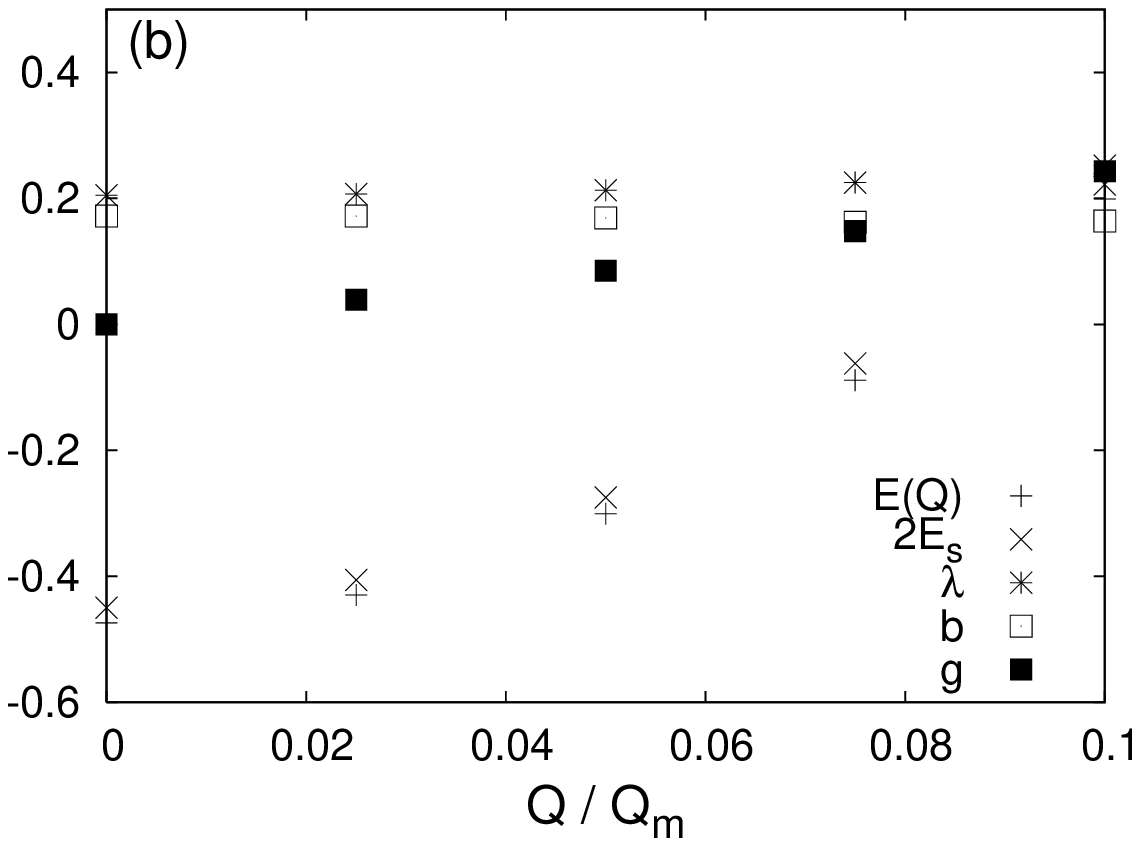}
    \caption{\label{fig:fig5}
      As for Fig.~\ref{fig:fig1}, but with $V^2=1$, $a=0.5$, and (a) $P=0$, (b) $P=1$.
    }
  \end{center}
\end{figure}

Figures~\ref{fig:fig5}(a) and~(b) show results of similar calculations as a function of $Q$
for a single value of $V^2$ for a more limited range of $Q$ for the
much smaller value of $a/r_\text{p}=0.5$ which may be appropriate for a quantum
wire of crystalline material.

The figures also show the values of $g$ of Eq.~(\ref{eq:g}) and of the
parameters $\lambda$ and $b$ which are found in the numerical work.  
We notice that in all cases
shown in Figs.~\ref{fig:fig1}\,--\,\ref{fig:fig5}, there is a monotonic rise
of energies with $Q$.

We have also considered two other types of trial functions for relative
motion, a Gaussian type of function with a second parameter in the
prefactor, and a wave function constant in $k$-space for wave vectors
with magnitudes between minimum and maximum values $k_1$ and $k_2$.
Both these types of trial function gave poorer results for the bipolaron
energies than the function used here.  Also, in many cases $k_1$ turned
out to be zero in the second type of function, and so our second
parameter often did not improve matters.

%
%
%
\section{\label{sec:discussion}Discussion}
%
%
%

Bipolaron energies measured from the bottom of the bare bands for the
Holstein model obtained from unpublished results of S. El Shawish for
the case of $\hbar\omega/t=16$, $g^2=0.125$ and $\hbar\omega/t=4$,
$g^2=0.5$, corresponding to $a=4$, $V^2=0.125$ and $a=2$, $V^2=0.5$ in
our model, are equal to -$0.335\hbar\omega$ and -$1.22\hbar\omega$
compared with -$0.29\hbar\omega$  and -$0.92\hbar\omega$ obtained by
us, and so our results are 13\% and 25\% smaller than those of El
Shawish for the magnitudes of this quantity.  The energies of two
single polarons in our model for the same parameters are
-$0.21\hbar\omega$ and -$0.64\hbar\omega$, and so our bipolaron
binding energies are $0.08\hbar\omega$ and $0.28\hbar\omega$.
Average effective masses (strictly the reciprocals of the average
reciprocal masses) of bipolarons up to $Q=0.1Q_m$ in our model for the
same pairs of parameters are $2.13m_\text{b}$ and $2.69m_\text{b}$,
whereas the average masses over the same region of wave vectors
inferred from the results of El Shawish are $2.88m_\text{b}$ and
$3.05m_\text{b}$.  Thus the departure of the mass from twice the
single-particle mass is much larger in the results of El Shawish. For
$a=4$, with $V^2=0.25$ and 0.375, bipolaron binding energies in our
model are $0.22 \hbar\omega$ and $0.38\hbar\omega$ respectively, while
bipolaron masses for these two cases are $2.31 m_b$ and $2.50 m_b$.
Thus we are able to get bipolaron binding energies of over a third
of the phonon energy without very large effective masses.

In the limit that $a\rightarrow 0$ (with $ta^2$ constant), our method
gives single-polaron binding energies $E_\text{s}=0.5 V^2a\hbar \omega$
and effective masses $m_\text{s}$  such that
$m_\text{s}/m_\text{b}=1+0.25V^2a$.  These results may be compared with
those of Chen \etal\cite{ChReJi98} for a continuum model with
short-range interactions with a single dispersionless phonon for values
of $V^2a$ up to three (corresponding to $\alpha$ up to 1.5 in their
notation).  For $V^2a=1$ ($\alpha=0.5$) and $V^2a=3$ ($\alpha=1.5$),
from Figs.~5 and 8 of their paper we find $E_\text{s}\approx 0.52\hbar
\omega$ and $E_\text{s}\approx 1.69\hbar \omega$, while their masses
$m_\text{s}$ satisfy $m_\text{s}/m_\text{b}\approx 1.2$ and 3.7 for the
two values of coupling.  Thus $E_\text{s}$ for the two models differ by
only 13\% for $V^2a=3$, whereas masses for the two models are within
4\% of each other for $V^2a=1$ (closer to 20\% for the departure of
$m/m_\text{b}$ from unity), but for $V^2a=3$ the discrepancy in masses
is large ($m_\text{s}/m_\text{b}=3.7$ in the model of
Ref.~\onlinecite{ChReJi98}, but 1.75 for our model).  The larger
discrepancy for masses between the two models than for binding energies
is related to the fact that intermediate-coupling methods are accurate
for binding energies up to higher values of coupling constants $\alpha$
in the Fr\"{o}hlich model than for masses.\cite{LeFlPi53,Sc59}

Another case for which we may make comparison with other authors is
that of $a=1$, $V^2=1$, corresponding to $\omega/t=1$  and $g^2=1$ in
the Hubbard-Holstein model.  Here we find that the bipolaron energy
measured from the bottom of the bare band is $-1.10\hbar \omega$, and
the energy of two single polarons is $-0.80\hbar\omega$.  Thus the
bipolaron binding energy with respect to two single polarons is
$0.30\hbar \omega$ in our model, compared with about $0.5\hbar \omega$
which can be inferred from the inset to Fig.~3 of a paper by Bon\v{c}a \etal\cite{BoKaTr00}  For this case we have also found the value of $Pa$ in our
model for which the bipolaron binding energies for the above values of
parameters vanish.  We find the bipolaron energy vanishes for
$Pa\approx 1.9$.  The result of a vanishing of the binding energy at
$Pa\approx 1.9$ may be compared with $U\approx 1.6$ for the binding
energy to vanish which may be inferred from the inset to Fig.~3 of
Ref.~\onlinecite{BoKaTr00}.

The smaller binding energies and masses, and larger partial bandwidths
cf. what are probably accurate results for the Hubbard-Holstein model
are thought to be due to a combination of (i) limitations of our
variational method to quite weak couplings, (ii) the use of only a
two-parameter model for the relative motion, 
and (iii) to differences between a Holstein
model and a model with constant bare mass.  We do not know which of the
three causes of discrepancy is most important at present.

A lower limit for the percentage change due to departure from weak
coupling of the expectation value of the part of the Hamiltonian which
does not depend on relative motion of the two electrons can be obtained
by looking at the percentage change in single-polaron energies at the
same value of coupling when this is known.  An approximate upper limit
to the percentage change in the same quantity can be obtained  from the
percentage change in the single-polaron binding energy when the
coupling $g^2$ is twice  as large, since, in the limit when the relative
wave function is very small, the coupling to phonons is twice as large
for a bipolaron as for a polaron.  For example, for the case $g=1$,
$t/\hbar \omega =1$, in the Holstein model, corresponding to $V^2=1$
and $a=1$ in our model, one can estimate from Fig.~1 of
Ref.~\onlinecite{RoBrLi99I} that
both the lower and upper limits in the errors in the bipolaron
energy below the bottom of the bare band due to use of a weak-coupling
method are very small.  

A simple way to estimate errors due to the change from the constant
bare-mass model to the Holstein model is to compare the weak-coupling
result for the single-polaron model in the Holstein model given in
Eq.~(8) of Ref.~\onlinecite{RoBrLi99I}, which gives a polaron binding energy $E_\text{bs}$ of
$g^2\hbar\omega/ (1+4J/\hbar\omega)^{\frac{1}{2}}$ in their notation.\cite{n3}
In our notation, $J/\hbar\omega$ corresponds to $1/a^2$, and $g$ to
$V$.  Thus their result in our notation is
\begin{equation}\label{eq:E_bs}
E_\text{bs}=V^2 a \hbar\omega/(4+a^2)^{\frac{1}{2}}.   
\end{equation}
Our method gives binding energies in units of $\hbar\omega$ for single
polarons in the constant bare-mass model of $0.21 V^2a$, $0.32 V^2a$
and $0.40 V^2a$ for $a$=4, 2 and 1, which may be compared with $0.22
V^2a$, $0.35 V^2a$ and $0.45 V^2a$ for the Holstein model  for weak
coupling from Eq.~(\ref{eq:E_bs}).  Thus the fractional changes in
binding energies at weak coupling due to use of the Holstein model
appear to be approximately 0.05, 0.09 and 0.12 for the three values of
$a$ considered.  The results presented above are for single-polaron
theory, and what changes would be expected for bipolarons is not
obvious.  However, if we take the results based on single-polaron
theory seriously for bipolarons, then we expect, \eg for $a=4$,
$V^2=0.125$, that the magnitude of the difference in the energy of the
bottom of the bare and bipolaron bands to be increased from about 
0.29$\hbar\omega$ for the constant bare-mass model to between 
0.30 and 0.31$\hbar
\omega$ for the Holstein model, compared with $0.335\hbar\omega$
obtained by El Shawish.  Thus in this case it appears probable that
errors due to the restriction of our method to weak or intermediate
coupling and due to use of a two-parameter trial function for relative
motion may be of the order of 10\% for bipolaron energies at $Q=0$.

Besides differences from the Hubbard model, we also discuss briefly how
far our method is likely to be fairly accurate within the framework of
the constant-bare-mass model that we have used.  We note the following
points:  

1. From the results mentioned after Eq.~(\ref{eq:E_bs}) and values of
parameters used for the figures and other values mentioned in the first
paragraph of this section, the single-polaron binding energy lies
between about 0.105$\hbar\omega$ (for $V^2=0.125$ and $a=4$) and
0.315$\hbar\omega$ (for $V^2=0.375$ and $a=4$).  The type of
variational method we use is accurate to about 4\% for single-polaron
binding energies in the Fr\"{o}hlich model and to within 20\% for
masses\cite{Sc59} up to polaron couping constants $\alpha=3$,
corresponding to polaron binding energies of about 3$\hbar\omega$.  Our
binding energies are so far below this value that our method can be
expected to be accurate for single polarons, and also for bipolarons if
one assumes that the coupling constant for bipolarons to have the same
percentage errors in masses as for single polarons is at least half as
large as that for single polarons. (See point 3 below for further
discussion of this). 

2. There is also the question of whether we are anywhere near a
transition between large and small polarons.  Toyozawa\cite{To61} was
the first to study such transitions, for the case of interactions
between electrons and acoustical phonons, and found a sudden
transition.  However, since we are dealing with optical phonons here,
his work does not have much relevance for our problem.  For
interactions with optical phonons such transitions (between large and
nearly-small polarons) were first discussed by one of the present
authors.\cite{Ea69,n2}

Emin\cite{Em73} considers transitions between large and small polarons
for a 3-D molecular-crystal model.  For the case where the bare half
bandwidth is ten times as large as the phonon energy, 
he finds a transition between types for the lowest-energy state as a
function of the binding energy of a small polaron which would exist in
the case of zero bare bandwidth.  He calls this energy $E_b$.  From
his Fig.2 one finds that the transition at $T=0$ for the above value of
bare half-bandwidth occurs when the polaron binding energy (not $E_b$)
is about $2\hbar\omega$, although both types of solutions exist for
weak-coupling polaron binding energies between about $\hbar\omega$ and
3$\hbar\omega$.  Even the lower bound on these values is considerably
larger than the values of polaron binding energies of up to
0.315$\hbar\omega$ which we have discussed in the present paper.  Thus
it is probable that our results are not influenced by any proximity to
such a transition.  However, Emin does not discuss in detail results in
the opposite limit which we have considered when bare band half
bandwidths are small compared with phonon energies.

For a model with Fr\"ohlich electron-phonon interactions, it appears
that a large to small polaron transition occurs for coupling constants
$\alpha$ near 3 to 5,\cite{Le92,Ia98} with details depending on the
degree of adiabaticity.  The transition is fairly sharp for
bare half bandwidths greater than phonon energies, but more gradual for the
opposite case.  For the smallest bare half bandwidth of $\hbar\omega$
considered in Ref.\onlinecite{Le92}, their Fig.3 indicates only a
fairly small departure of masses from a linear dependence on coupling
constant $\alpha$ up to one.  The cases we concentrate on are  $a=4$
and $a=2$, corresponding to bare half bandwidths of 0.31$\hbar\omega$
and $1.23\hbar\omega$ for a constant bare mass up to the edge of the
Brillouin zone at $\pi/a$, or to 0.125$\hbar\omega$ and 0.5$\hbar\omega$
if we convert masses to obtain transfer integrals $t$ for a
tight-binding model with the same bare mass at the bottom of the band.
Thus, although our coupling constants correspond to polaron binding
energies considerably smaller than $\hbar\omega$ in the Fr\"ohlich
model, we could not completely rule out a larger mass rise than we find
by our method because of the beginning of a transition between polaron
types if our band structure were similar to that of a tight-binding
model as considered in Refs.\onlinecite{Le92,Ia98}.  However, in our
case we are assuming that any effect of the lattice periodicity on the
bare electrons is small, and so in such a case any transitions between
polaron types will correspond to transitions in the continuum model,
i.e. to a change from the lattice following the instantaneous position
of the electron for weak coupling to responding to some average position
for stronger coupling, and the effects of such transitions are already
included in calculations such as those reported in
Ref.\onlinecite{Sc59}.  

For our case where there is no significant effect of periodicity of
the potential on the bare-electron wave function, there is no such thing
as a small polaron in the sense of a state which is a linear
combination of states with the electron on one lattice site and
surrounded by the appropriate lattice polarisation, since lattice sites
are almost indistinguishable from positions in between them.  In this
case the fact that our single-polaron radius for $a=4$ is only a
quarter of the lattice constant does not imply small polarons in the
usual sense.  The only way that the lattice constant comes into our
model is by a cut-off in the phonon wave vector.  Thus in this sense it
is similar to the continuum-polarisation model with a cut-off
considered by Schultz.\cite{Sc59}

3. Iadonisi et al.\cite{IaPeCadF01} also consider transitions between
types for bipolarons.  For a case shown in Fig.1 of their paper,
corresponding to a bare bandwidth equal to twenty times the phonon
energy, the transition for bipolarons occurs at about an 8\% smaller
values of the coupling constant than for polarons.  Thus our guess
in point 1 above that a given percentage error of mass may occur
at a value of coupling constant of about half that of polarons may be
pessimistic.

We get no confirmation in this work of our conjectures based on
perturbation theory of great enhancements of pair binding energies at
certain large centre-of-mass wave vectors.\cite{Ea94,Ea94_2} Also,
unpublished calculations of El Shawish up to bipolaron wave vectors of
$\pi/a$ do not give us much reason to expect that suggestions of a
cusp-like minimum at $2\pi/a$ (in an extended zone scheme) indicated by
early attempts to extend results of our variational method to the
Hubbard-Holstein model as reported in Ref.~\onlinecite{Ea99}, are
likely to occur in accurate calculations.  However, we still cannot
rule out the possibility that a dip in bipolaron energies would be
obtained near certain wave vectors if we were to use a different type
of variational wave function for relative motion which could take
better advantage of the small denominators in the integrands in
Eq.~(\ref{eq:E(Q)}) for suitable values of $Q$ and fairly weak
coupling.

Our calculations indicate that there are parameter values where
bipolarons do not have excessively high masses while having binding
energies with respect to two single-polaron energies greater
than a few tenths of the relevant boson energy.  If the bosons are
plasmons of energy of the order of 2 eV, then this permits bipolaron binding
energies of the order of 0.5 eV without too great increases in masses,
whereas for phonons of energies of about 0.36 eV, bipolaron binding
energies of 0.1 eV can be obtained without too large mass increases.  A
binding energy of at least 0.1 eV is a minimum requirement for
room-temperature superconductivity, assuming pair binding energies must
be at least about $4k_\text{B}T$ for superconductivity at temperature $T$.
The masses must not be too high in order to be able to have a high
Bose-Einstein condensation temperature for bipolarons without excessively
high bipolaron concentrations.  Previous calculations of condensation
temperatures for bosons with a quadratic $E(Q)$ curve\cite{Ea98} in arrays of
nanofilaments have recently been extended\cite{Ea05} to cases with a
dispersion approximated by a sum of linear and quadratic terms, as
indicated to occur for Cooper pairs.\cite{AdCaPuRiFoSoDLVaRo00} We hope to modify the
calculations of Ref.~\onlinecite{Ea05} soon by use of a Bogoliubov-type of
dispersion for pairs (see \eg Ref.~\onlinecite{PiSt05}), which we now think
is more appropriate than that based on a Cooper-pair model for the strongly
coupled pairs at fairly low carrier concentrations in which we were
interested in Ref.~\onlinecite{Ea05}.

%
%
%
\section{\label{sec:conclusions}Conclusions}
%
%
%

No support comes from our variational method for previous results based
on perturbation theory that great enhancements of binding energies of
pairs can be obtained at appropriate large centre-of-mass velocities.
However, parameters have been found such that bipolaron masses are
smaller than about $3m_\text{e}$ while keeping binding energies with respect to
energies of two single polarons greater than 0.1 eV.  Thus bipolarons
in one dimension may provide a basis for an explanation of probable
room-temperature superconductivity in narrow channels through films of
oxidised atactic polypropylene and other polymers, but for different
reasons than conjectured in earlier papers.

\begin{acknowledgments}
  
  One of us (DME) would like to thank Professor J. Robles-Dominguez for
  weekly discussions over a period of a few months on bipolarons and
  related subjects, Dr. S. El Shawish for sending results on
  bipolaron dispersion of the Holstein model for some parameters of
  interest, and Dr. M. Hohenadler for help in preparation of the manuscript.

\end{acknowledgments}



\begin{thebibliography}{65}
\expandafter\ifx\csname natexlab\endcsname\relax\def\natexlab#1{#1}\fi
\expandafter\ifx\csname bibnamefont\endcsname\relax
  \def\bibnamefont#1{#1}\fi
\expandafter\ifx\csname bibfnamefont\endcsname\relax
  \def\bibfnamefont#1{#1}\fi
\expandafter\ifx\csname citenamefont\endcsname\relax
  \def\citenamefont#1{#1}\fi
\expandafter\ifx\csname url\endcsname\relax
  \def\url#1{\texttt{#1}}\fi
\expandafter\ifx\csname urlprefix\endcsname\relax\def\urlprefix{URL }\fi
\providecommand{\bibinfo}[2]{#2}
\providecommand{\eprint}[2][]{\url{#2}}

\bibitem[{\citenamefont{Vinetskii}(1961)}]{Vi61}
\bibinfo{author}{\bibfnamefont{V.~L.} \bibnamefont{Vinetskii}},
  \bibinfo{journal}{Zh. Eksp. Teor. Fiz.} \textbf{\bibinfo{volume}{40}},
  \bibinfo{pages}{1459} (\bibinfo{year}{1961}), \bibinfo{note}{[Sov. Phys.
  JETP, 13 (1961) 1023]}.

\bibitem[{\citenamefont{Hiramoto and Toyozawa}(1985)}]{HiTo85}
\bibinfo{author}{\bibfnamefont{H.}~\bibnamefont{Hiramoto}} \bibnamefont{and}
  \bibinfo{author}{\bibfnamefont{Y.}~\bibnamefont{Toyozawa}},
  \bibinfo{journal}{J. Phys. Soc. Jpn.} \textbf{\bibinfo{volume}{54}},
  \bibinfo{pages}{245} (\bibinfo{year}{1985}).

\bibitem[{\citenamefont{Adamowski}(1989)}]{Ad89}
\bibinfo{author}{\bibfnamefont{J.}~\bibnamefont{Adamowski}},
  \bibinfo{journal}{Phys. Rev. B} \textbf{\bibinfo{volume}{39}},
  \bibinfo{pages}{3649} (\bibinfo{year}{1989}).

\bibitem[{\citenamefont{G~Verbist and Devreese}(1991)}]{VePeDe91}
\bibinfo{author}{\bibfnamefont{G.} \bibnamefont{Verbist}},
\bibinfo{author}{\bibfnamefont{F.~M.} \bibnamefont{Peeters}},
  \bibnamefont{and} \bibinfo{author}{\bibfnamefont{J.~T.}
  \bibnamefont{Devreese}}, \bibinfo{journal}{Phys. Rev. B}
  \textbf{\bibinfo{volume}{43}}, \bibinfo{pages}{2712} (\bibinfo{year}{1991}).

\bibitem[{\citenamefont{Bassani et~al.}(1991)\citenamefont{Bassani, Geddo,
  Iadonisi, and Ninno}}]{BaGeIaNi91}
\bibinfo{author}{\bibfnamefont{F.}~\bibnamefont{Bassani}},
  \bibinfo{author}{\bibfnamefont{M.}~\bibnamefont{Geddo}},
  \bibinfo{author}{\bibfnamefont{G.}~\bibnamefont{Iadonisi}}, \bibnamefont{and}
  \bibinfo{author}{\bibfnamefont{D.}~\bibnamefont{Ninno}},
  \bibinfo{journal}{Phys. Rev. B} \textbf{\bibinfo{volume}{43}},
  \bibinfo{pages}{5296} (\bibinfo{year}{1991}).

\bibitem[{\citenamefont{Emin and Hillery}(1989)}]{EmHi89}
\bibinfo{author}{\bibfnamefont{D.}~\bibnamefont{Emin}} \bibnamefont{and}
  \bibinfo{author}{\bibfnamefont{M.~S.} \bibnamefont{Hillery}},
  \bibinfo{journal}{Phys. Rev. B} \textbf{\bibinfo{volume}{39}},
  \bibinfo{pages}{6575} (\bibinfo{year}{1989}).

\bibitem[{\citenamefont{Dzhumanov et~al.}(1996)\citenamefont{Dzhumanov,
  Baratov, and Abboudy}}]{DzBaAb96}
\bibinfo{author}{\bibfnamefont{S.}~\bibnamefont{Dzhumanov}},
  \bibinfo{author}{\bibfnamefont{A.~A.}~\bibnamefont{Baratov}}, \bibnamefont{and}
  \bibinfo{author}{\bibfnamefont{S.}~\bibnamefont{Abboudy}}, 
  \bibinfo{journal}{Phys. Rev. B} \textbf{\bibinfo{volume}{54}},
  \bibinfo{pages}{13121} (\bibinfo{year}{1996}).

\bibitem[{\citenamefont{Ba\u{i}matov et~al.}(1997)\citenamefont{Ba\u{i}matov,
  Khuzhakulov, and Sharipov}}]{BaKhSh97}
\bibinfo{author}{\bibfnamefont{P.~Zh.}~\bibnamefont{Ba\u{i}matov}},
  \bibinfo{author}{\bibfnamefont{D.~Ch.}~\bibnamefont{Khuzhakulov}},
  \bibnamefont{and} \bibinfo{author}{\bibfnamefont{Kh.~T.}
  \bibnamefont{Sharipov}}, \bibinfo{journal}{Fiz. Tverd. Tela}
  \textbf{\bibinfo{volume}{39}}, \bibinfo{pages}{284} (\bibinfo{year}{1997})
  \bibinfo{note}{[Phys. Solid State {\bf 39}, 248 (1997)]}.

\bibitem[{\citenamefont{da~Costa and Peeters}(1998)}]{dCPe98}
\bibinfo{author}{\bibfnamefont{W.~B.}~\bibnamefont{da~Costa}} \bibnamefont{and}
  \bibinfo{author}{\bibfnamefont{F.~M.} \bibnamefont{Peeters}},
  \bibinfo{journal}{Phys. Rev. B} \textbf{\bibinfo{volume}{57}},
  \bibinfo{pages}{10569} (\bibinfo{year}{1998}).

\bibitem[{\citenamefont{Pokatilov et~al.}(2000)\citenamefont{Pokatilov, Fomin,
  Devreese, Balaban, and Klimin}}]{PoFoDeBaKl00}
\bibinfo{author}{\bibfnamefont{E.~B.} \bibnamefont{Pokatilov}},
  \bibinfo{author}{\bibfnamefont{V.~M.} \bibnamefont{Fomin}},
  \bibinfo{author}{\bibfnamefont{J.~T.} \bibnamefont{Devreese}},
  \bibinfo{author}{\bibfnamefont{S.~N.} \bibnamefont{Balaban}},
  \bibnamefont{and} \bibinfo{author}{\bibfnamefont{S.~M.}
  \bibnamefont{Klimin}}, \bibinfo{journal}{Phys. Rev. B}
  \textbf{\bibinfo{volume}{61}}, \bibinfo{pages}{2721} (\bibinfo{year}{2000}),
  \bibinfo{note}{and references therein}.

\bibitem[{\citenamefont{Myasnikova}(2001)}]{My01}
\bibinfo{author}{\bibfnamefont{A.~E.} \bibnamefont{Myasnikova}},
  \bibinfo{journal}{Phys. Lett. A}  
  \textbf{\bibinfo{volume}{291}}, \bibinfo{pages}{439} 
(\bibinfo{year}{2001}).

\bibitem[{\citenamefont{Mukhomorov}(2002)}]{Mu02}
\bibinfo{author}{\bibfnamefont{V.~K.} \bibnamefont{Mukhomorov}},
  \bibinfo{journal}{Fiz. Tverd. Tela} \textbf{\bibinfo{volume}{44}},
  \bibinfo{pages}{232} (\bibinfo{year}{2002}), \bibinfo{note}{[Phys. Solid
  State {\bf 44}, 241 (2002)]};
\bibinfo{journal}{Fiz. Tverd. Tela}
  \textbf{\bibinfo{volume}{48}}, \bibinfo{pages}{622}
  (\bibinfo{year}{2006}) (\bibinfo{note}{[Phys. Solid State {\bf 46}, 864 (2006)
]}. (\bibinfo{note}{Kashyrina et al., in a 2003 paper,
Ref.~\onlinecite{KaLaSy03}, discuss what they report to be errors in
several papers by Mukhomorov.)}

\bibitem[{\citenamefont{Senger and Er\c{c}elebi}(2002)}]{SeEr02}
\bibinfo{author}{\bibfnamefont{R.~T.} \bibnamefont{Senger}} \bibnamefont{and}
  \bibinfo{author}{\bibfnamefont{A.}~\bibnamefont{Er\c{c}elebi}},
  \bibinfo{journal}{Eur. Phys. J. B} \textbf{\bibinfo{volume}{26}},
  \bibinfo{pages}{253} (\bibinfo{year}{2002}).

\bibitem[{\citenamefont{Kandemir}(2004)}]{Ka04}
\bibinfo{author}{\bibfnamefont{B.~S.} \bibnamefont{Kandemir}},
  \bibinfo{journal}{Eur. Phys. J. B} \textbf{\bibinfo{volume}{37}},
  \bibinfo{pages}{527} (\bibinfo{year}{2004});
\bibinfo{journal}{Phys. Rev. B}
  \textbf{\bibinfo{volume}{74}}, \bibinfo{pages}{075330}
  (\bibinfo{year}{2006}). 

\bibitem[{\citenamefont{Kashyrina et~al.}(2003)\citenamefont{Kashyrina, Lakhno,
  and Sychkov}}]{KaLaSy03}
\bibinfo{author}{\bibfnamefont{N.~I.} \bibnamefont{Kashyrina}},
  \bibinfo{author}{\bibfnamefont{V.~D.} \bibnamefont{Lakhno}},
  \bibnamefont{and} \bibinfo{author}{\bibfnamefont{V.~V.}
  \bibnamefont{Sychyov}}, 
\bibinfo{journal}{phys. stat. sol. (b)}
  \textbf{\bibinfo{volume}{239}}, \bibinfo{pages}{174}
  (\bibinfo{year}{2003});
\bibinfo{journal}{Phys. Rev. B}
  \textbf{\bibinfo{volume}{71}}, \bibinfo{pages}{134301}
  (\bibinfo{year}{2005}).

\bibitem[{\citenamefont{Chakraverty and Schlenker}(1976)}]{ChSc76}
\bibinfo{author}{\bibfnamefont{B.~K.} \bibnamefont{Chakraverty}}
  \bibnamefont{and}
  \bibinfo{author}{\bibfnamefont{C.}~\bibnamefont{Schlenker}},
  \bibinfo{journal}{J. Physique}  
  \textbf{\bibinfo{volume}{37}}, \bibinfo{pages}{Colloq.C4-353}
(\bibinfo{year}{1976}).

\bibitem[{\citenamefont{Alexandrov and Ranninger}(1981)}]{AlRa81}
\bibinfo{author}{\bibfnamefont{A.~S.} \bibnamefont{Alexandrov}}
  \bibnamefont{and}
  \bibinfo{author}{\bibfnamefont{J.}~\bibnamefont{Ranninger}},
  \bibinfo{journal}{Phys. Rev. B} \textbf{\bibinfo{volume}{23}},
  \bibinfo{pages}{1796} (\bibinfo{year}{1981}).

\bibitem[{\citenamefont{Cohen et~al.}(1984)\citenamefont{Cohen, Economou
  and Soukoulis}}]{CoEcSo84}
\bibinfo{author}{\bibfnamefont{M.~H.} \bibnamefont{Cohen}},
  \bibinfo{author}{\bibfnamefont{E.~N.} \bibnamefont{Economou}},
  \bibnamefont{and} \bibinfo{author}{\bibfnamefont{C.~M.}
  \bibnamefont{Soukoulis}}, \bibinfo{journal}{Phys. Rev. B}
  \textbf{\bibinfo{volume}{29}}, \bibinfo{pages}{4496}
  (\bibinfo{year}{1984}).

\bibitem[{\citenamefont{Ranninger and Thibblin}(1992)}]{RaTh92}
\bibinfo{author}{\bibfnamefont{J.}~\bibnamefont{Ranninger}} \bibnamefont{and}
  \bibinfo{author}{\bibfnamefont{U.}~\bibnamefont{Thibblin}},
  \bibinfo{journal}{Phys. Rev. B} \textbf{\bibinfo{volume}{45}},
  \bibinfo{pages}{7730} (\bibinfo{year}{1992}).

\bibitem[{sma()}]{smallbipolarons}
\bibinfo{note}{For further early references on small bipolarons see: \newline
  D.~M. Eagles, Physica C {\bf 156}, 382 (1988); \newline A.~S. Alexandrov
  and N.~F. Mott, Polarons and Bipolarons (World Scientific, Singapore, 1995)}.

\bibitem[{\citenamefont{Alexandrov and Mott}(1994)}]{AlMo94}
\bibinfo{author}{\bibfnamefont{A.~S.} \bibnamefont{Alexandrov}}
  \bibnamefont{and} \bibinfo{author}{\bibfnamefont{N.~F.}
  \bibnamefont{Mott}}, \bibinfo{journal}{Rep. Prog. Phys.}
  \textbf{\bibinfo{volume}{57}}, \bibinfo{pages}{1197} (\bibinfo{year}{1994}).

\bibitem[{\citenamefont{Alexandrov}(1996)}]{Al96}
\bibinfo{author}{\bibfnamefont{A.~S.} \bibnamefont{Alexandrov}},
  \bibinfo{journal}{Phys. Rev. B} \textbf{\bibinfo{volume}{53}},
  \bibinfo{pages}{2863} (\bibinfo{year}{1996}).

\bibitem[{\citenamefont{Alexandrov and Bratkovsky}(1999)}]{AlBr99}
\bibinfo{author}{\bibfnamefont{A.~S.} \bibnamefont{Alexandrov}}
  \bibnamefont{and} \bibinfo{author}{\bibfnamefont{A.~M.}
  \bibnamefont{Bratkovsky}}, \bibinfo{journal}{J. Phys.: Condens. Matter}
  \textbf{\bibinfo{volume}{11}}, \bibinfo{pages}{L531} (\bibinfo{year}{1999}).

\bibitem[{\citenamefont{Alexandrov and Kornilovitch}(2002)}]{AlKo02}
\bibinfo{author}{\bibfnamefont{A.~S.} \bibnamefont{Alexandrov}}
  \bibnamefont{and} \bibinfo{author}{\bibfnamefont{P.~E.}
  \bibnamefont{Kornilovitch}}, \bibinfo{journal}{J. Phys.: Condens. Matter}
  \textbf{\bibinfo{volume}{14}}, \bibinfo{pages}{5337} (\bibinfo{year}{2002}).

\bibitem[{\citenamefont{Marsiglio}(1995)}]{Marsiglio95}
\bibinfo{author}{\bibfnamefont{F.}~\bibnamefont{Marsiglio}},
  \bibinfo{journal}{Physica C} \textbf{\bibinfo{volume}{244}},
  \bibinfo{pages}{21} (\bibinfo{year}{1995}).

\bibitem[{\citenamefont{Fehske et~al.}(1995{\natexlab{a}})\citenamefont{Fehske,
  R\"oder, Wellein, and Mistriotis}}]{FeRoWeMi95}
\bibinfo{author}{\bibfnamefont{H.}~\bibnamefont{Fehske}},
  \bibinfo{author}{\bibfnamefont{H.}~\bibnamefont{R\"oder}},
  \bibinfo{author}{\bibfnamefont{G.}~\bibnamefont{Wellein}}, \bibnamefont{and}
  \bibinfo{author}{\bibfnamefont{A.}~\bibnamefont{Mistriotis}},
  \bibinfo{journal}{Phys. Rev. B} \textbf{\bibinfo{volume}{51}},
  \bibinfo{pages}{16 582} (\bibinfo{year}{1995}{\natexlab{a}}).

\bibitem[{\citenamefont{Wellein et~al.}(1996)\citenamefont{Wellein, R\"oder,
  and Fehske}}]{WeRoFe96}
\bibinfo{author}{\bibfnamefont{G.}~\bibnamefont{Wellein}},
  \bibinfo{author}{\bibfnamefont{H.}~\bibnamefont{R\"oder}}, \bibnamefont{and}
  \bibinfo{author}{\bibfnamefont{H.}~\bibnamefont{Fehske}},
  \bibinfo{journal}{Phys. Rev. B} \textbf{\bibinfo{volume}{53}},
  \bibinfo{pages}{9666} (\bibinfo{year}{1996}).

\bibitem[{\citenamefont{Fehske et~al.}(2000)\citenamefont{Fehske, Loos, and
  Wellein}}]{FeLoWe00}
\bibinfo{author}{\bibfnamefont{H.}~\bibnamefont{Fehske}},
  \bibinfo{author}{\bibfnamefont{J.}~\bibnamefont{Loos}}, \bibnamefont{and}
  \bibinfo{author}{\bibfnamefont{G.}~\bibnamefont{Wellein}},
  \bibinfo{journal}{Phys. Rev. B} \textbf{\bibinfo{volume}{61}},
  \bibinfo{pages}{8016} (\bibinfo{year}{2000}).

\bibitem[{\citenamefont{{La Magna} and Pucci}(1997)}]{LaMaPu97}
\bibinfo{author}{\bibfnamefont{A.}~\bibnamefont{{La Magna}}} \bibnamefont{and}
  \bibinfo{author}{\bibfnamefont{R.}~\bibnamefont{Pucci}},
  \bibinfo{journal}{Phys. Rev. B} \textbf{\bibinfo{volume}{55}},
  \bibinfo{pages}{14 886} (\bibinfo{year}{1997}).

\bibitem[{\citenamefont{Firsov and Kudinov}(1997)}]{FiKu97}
\bibinfo{author}{\bibfnamefont{Y.~A.} \bibnamefont{Firsov}} \bibnamefont{and}
  \bibinfo{author}{\bibfnamefont{E.~K.} \bibnamefont{Kudinov}},
  \bibinfo{journal}{Phys. Solid State} \textbf{\bibinfo{volume}{39}},
  \bibinfo{pages}{1930} (\bibinfo{year}{1997}).

\bibitem[{\citenamefont{Proville and Aubry}(1999)}]{PrAu99}
\bibinfo{author}{\bibfnamefont{L.}~\bibnamefont{Proville}} \bibnamefont{and}
  \bibinfo{author}{\bibfnamefont{S.}~\bibnamefont{Aubry}},
  \bibinfo{journal}{Eur. Phys. J. B} \textbf{\bibinfo{volume}{11}},
  \bibinfo{pages}{41} (\bibinfo{year}{1999}).

\bibitem[{\citenamefont{Frank and Wagner}(1999)}]{FrWa99}
\bibinfo{author}{\bibfnamefont{Th.}~\bibnamefont{Frank}} \bibnamefont{and}
  \bibinfo{author}{\bibfnamefont{M.}~\bibnamefont{Wagner}},
  \bibinfo{journal}{Phys. Rev. B} \textbf{\bibinfo{volume}{60}},
  \bibinfo{pages}{3252} (\bibinfo{year}{1999}).

\bibitem[{\citenamefont{Sil}(1999)}]{Sil99}
\bibinfo{author}{\bibfnamefont{S.}~\bibnamefont{Sil}}, \bibinfo{journal}{J.
  Phys.: Condens. Matter} \textbf{\bibinfo{volume}{11}}, \bibinfo{pages}{8879}
  (\bibinfo{year}{1999}).

\bibitem[{\citenamefont{Zhang et~al.}(1999)\citenamefont{Zhang, Jeckelman
  and White}}]{ZhJeWh99}
\bibinfo{author}{\bibfnamefont{C.} \bibnamefont{Zhang}},
  \bibinfo{author}{\bibfnamefont{E.} \bibnamefont{Jeckelmann}},
  \bibnamefont{and} \bibinfo{author}{\bibfnamefont{S.~R.}
  \bibnamefont{White}}, \bibinfo{journal}{Phys. Rev. B}
  \textbf{\bibinfo{volume}{60}}, \bibinfo{pages}{14092}
  (\bibinfo{year}{1999}).

\bibitem[{\citenamefont{Bon\v{c}a et~al.}(2000)\citenamefont{Bon\v{c}a,
  Katra\v{s}nik, and Trugman}}]{BoKaTr00}
\bibinfo{author}{\bibfnamefont{J.}~\bibnamefont{Bon\v{c}a}},
  \bibinfo{author}{\bibfnamefont{T.}~\bibnamefont{Katra\v{s}nik}},
  \bibnamefont{and} \bibinfo{author}{\bibfnamefont{S.~A.}
  \bibnamefont{Trugman}}, \bibinfo{journal}{Phys. Rev. Lett.}
  \textbf{\bibinfo{volume}{84}}, \bibinfo{pages}{3153} (\bibinfo{year}{2000}).

\bibitem[{\citenamefont{Bon\v{c}a and Trugman}(2001)}]{BoTr01}
\bibinfo{author}{\bibfnamefont{J.}~\bibnamefont{Bon\v{c}a}} \bibnamefont{and}
  \bibinfo{author}{\bibfnamefont{S.~A.} \bibnamefont{Trugman}},
  \bibinfo{journal}{Phys. Rev. B} \textbf{\bibinfo{volume}{64}},
  \bibinfo{pages}{094507} (\bibinfo{year}{2001}).

\bibitem[{\citenamefont{{El Shawish} et~al.}(2003)\citenamefont{{El Shawish},
  Bon\v{c}a, Ku, and Trugman}}]{ElShBoKuTr03}
\bibinfo{author}{\bibfnamefont{S.}~\bibnamefont{{El Shawish}}},
  \bibinfo{author}{\bibfnamefont{J.}~\bibnamefont{Bon\v{c}a}},
  \bibinfo{author}{\bibfnamefont{L.~C.} \bibnamefont{Ku}}, \bibnamefont{and}
  \bibinfo{author}{\bibfnamefont{S.~A.} \bibnamefont{Trugman}},
  \bibinfo{journal}{Phys. Rev. B} \textbf{\bibinfo{volume}{67}},
  \bibinfo{pages}{014301} (\bibinfo{year}{2003}).

\bibitem[{\citenamefont{Iadonisi et~al.}(2001)\citenamefont{Iadonisi, Perroni,
  Cataudella, and {de Filippis}}}]{IaPeCadF01}
\bibinfo{author}{\bibfnamefont{G.}~\bibnamefont{Iadonisi}},
  \bibinfo{author}{\bibfnamefont{C.~A.} \bibnamefont{Perroni}},
  \bibinfo{author}{\bibfnamefont{V.}~\bibnamefont{Cataudella}},
  \bibnamefont{and} \bibinfo{author}{\bibfnamefont{G.}~\bibnamefont{{de
  Filippis}}}, \bibinfo{journal}{J. Phys.: Condens. Matter}
  \textbf{\bibinfo{volume}{13}}, \bibinfo{pages}{1499} (\bibinfo{year}{2001}).

\bibitem[{\citenamefont{{De Filippis} et~al.}(2001)\citenamefont{{De Filippis},
  Cataudella, Iadonisi, {Marigliano Ramaglia}, Perroni, and
  Ventriglia}}]{deFiCaIaMaPeVe01}
\bibinfo{author}{\bibfnamefont{G.}~\bibnamefont{{De Filippis}}},
  \bibinfo{author}{\bibfnamefont{V.}~\bibnamefont{Cataudella}},
  \bibinfo{author}{\bibfnamefont{G.}~\bibnamefont{Iadonisi}},
  \bibinfo{author}{\bibfnamefont{V.}~\bibnamefont{{Marigliano Ramaglia}}},
  \bibinfo{author}{\bibfnamefont{C.~A.} \bibnamefont{Perroni}},
  \bibnamefont{and}
  \bibinfo{author}{\bibfnamefont{F.}~\bibnamefont{Ventriglia}},
  \bibinfo{journal}{Phys. Rev. B} \textbf{\bibinfo{volume}{64}},
  \bibinfo{pages}{155105} (\bibinfo{year}{2001}).

\bibitem[{\citenamefont{{Macridin} et~al.}(2004)\citenamefont{Macridin,
 Sawatsky and Jarrell}}]{Ma04} 
\bibinfo{author}{\bibfnamefont{A.}~\bibnamefont{Macridin}},
  \bibinfo{author}{\bibfnamefont{G.~A.}~\bibnamefont{Sawatzky}},
  \bibnamefont{and}
  \bibinfo{author}{\bibfnamefont{M.}~\bibnamefont{Jarrell}},
  \bibinfo{journal}{Phys. Rev. B} \textbf{\bibinfo{volume}{69}},
  \bibinfo{pages}{245111} (\bibinfo{year}{2004}).

\bibitem[{\citenamefont{Hohenadler et~al.}(2005)\citenamefont{Hohenadler,
  Aichhorn, and {von der Linden}}}]{HoAivdL04}
\bibinfo{author}{\bibfnamefont{M.}~\bibnamefont{Hohenadler}},
  \bibinfo{author}{\bibfnamefont{M.}~\bibnamefont{Aichhorn}}, \bibnamefont{and}
  \bibinfo{author}{\bibfnamefont{W.}~\bibnamefont{{von der Linden}}},
  \bibinfo{journal}{Phys. Rev. B} \textbf{\bibinfo{volume}{71}},
  \bibinfo{pages}{014302} (\bibinfo{year}{2005}).

\bibitem[{\citenamefont{Hohenadler and von~der Linden}(2005)}]{HovdL05}
\bibinfo{author}{\bibfnamefont{M.}~\bibnamefont{Hohenadler}} \bibnamefont{and}
  \bibinfo{author}{\bibfnamefont{W.}~\bibnamefont{von~der Linden}},
  \bibinfo{journal}{Phys. Rev. B} \textbf{\bibinfo{volume}{71}},
  \bibinfo{pages}{184309} (\bibinfo{year}{2005}).


\bibitem[{\citenamefont{Zheng et~al.}(2005)}]
  {Zh05}
\bibinfo{author}{\bibfnamefont{L.~Y.}~\bibnamefont{Zheng}},
  \bibinfo{author}{\bibfnamefont{Y.~N.}~\bibnamefont{Chiu}},
  \bibnamefont{and} \bibinfo{author}{\bibfnamefont{S.~T.}~\bibnamefont
 {Lai}},
  \bibinfo{journal}{Journal of Molecular Structure (Theochem)} \textbf{\bibinfo{volume}{722}},
  \bibinfo{pages}{147} (\bibinfo{year}{2005}).


\bibitem[{\citenamefont{Gurari}(1953)}]{Gu53}
\bibinfo{author}{\bibfnamefont{M.}~\bibnamefont{Gurari}},
  \bibinfo{journal}{Phil. Mag.} \textbf{\bibinfo{volume}{44}},
  \bibinfo{pages}{329} (\bibinfo{year}{1953}).

\bibitem[{\citenamefont{Fr\"{o}hlich}(1954)}]{Fr54}
\bibinfo{author}{\bibfnamefont{H.}~\bibnamefont{Fr\"{o}hlich}},
  \bibinfo{journal}{Adv. Phys.} \textbf{\bibinfo{volume}{3}},
  \bibinfo{pages}{325} (\bibinfo{year}{1954}).

\bibitem[{\citenamefont{Lee and Pines}(1952)}]{LePi52}
\bibinfo{author}{\bibfnamefont{T.~D.} \bibnamefont{Lee}} \bibnamefont{and}
  \bibinfo{author}{\bibfnamefont{D.}~\bibnamefont{Pines}},
  \bibinfo{journal}{Phys. Rev.} \textbf{\bibinfo{volume}{88}},
  \bibinfo{pages}{960} (\bibinfo{year}{1952}).

\bibitem[{\citenamefont{Lee et~al.}(1953)\citenamefont{Lee, Low, and
  Pines}}]{LeFlPi53}
\bibinfo{author}{\bibfnamefont{T.~D.} \bibnamefont{Lee}},
  \bibinfo{author}{\bibfnamefont{F.}~\bibnamefont{Low}}, \bibnamefont{and}
  \bibinfo{author}{\bibfnamefont{D.}~\bibnamefont{Pines}},
  \bibinfo{journal}{Phys. Rev.} \textbf{\bibinfo{volume}{90}},
  \bibinfo{pages}{297} (\bibinfo{year}{1953}).

\bibitem[{\citenamefont{Hubbard}(1963)}]{Hu63}
\bibinfo{author}{\bibfnamefont{J.}~\bibnamefont{Hubbard}},
  \bibinfo{journal}{Proc. R. Soc. London} \textbf{\bibinfo{volume}{276}},
  \bibinfo{pages}{238} (\bibinfo{year}{1963}).

\bibitem[{\citenamefont{Holstein}(1959)}]{Ho59a}
\bibinfo{author}{\bibfnamefont{T.}~\bibnamefont{Holstein}},
  \bibinfo{journal}{Ann. Phys. (N.Y.)} \textbf{\bibinfo{volume}{8}},
  \bibinfo{pages}{325; {\bf 8}, 343} (\bibinfo{year}{1959}).

\bibitem[{\citenamefont{Parmenter}(1959)}]{Pa59}
\bibinfo{author}{\bibfnamefont{R.~H.} \bibnamefont{Parmenter}},
  \bibinfo{journal}{Phys. Rev.} \textbf{\bibinfo{volume}{116}},
  \bibinfo{pages}{1390} (\bibinfo{year}{1959}); \bibinfo{note}{ibid. {\bf 140},
  A1952 (1965)}.

\bibitem[{\citenamefont{Eagles}(1966)}]{Ea66}
\bibinfo{author}{\bibfnamefont{D.~M.} \bibnamefont{Eagles}},
  \bibinfo{journal}{Phys. Lett.} \textbf{\bibinfo{volume}{20}},
  \bibinfo{pages}{591} (\bibinfo{year}{1966}).

\bibitem[{\citenamefont{Eagles}(1994{\natexlab{a}})}]{Ea94}
\bibinfo{author}{\bibfnamefont{D.~M.} \bibnamefont{Eagles}},
  \bibinfo{journal}{Proc. Conf. Phys. Chem. Molecular and Oxide
  Superconductors, J. Supercond.} 
\textbf{\bibinfo{volume}{7}},
  \bibinfo{pages}{679} (\bibinfo{year}{1994}{\natexlab{a}}).

\bibitem[{\citenamefont{Eagles}(1994{\natexlab{b}})}]{Ea94_2}
\bibinfo{author}{\bibfnamefont{D.~M.} \bibnamefont{Eagles}},
  \bibinfo{journal}{Physica C}  
\textbf{\bibinfo{volume}{225}}, \bibinfo{pages}{222}
(\bibinfo{year}{1994}{\natexlab{b}}),
  \bibinfo{note}{erratum ibid. {\bf 280}, 335 (1997)}.

\bibitem[{\citenamefont{Hone}(1965)}]{Ho65}
\bibinfo{author}{\bibfnamefont{D.}~\bibnamefont{Hone}}, \bibinfo{journal}{Phys.
  Rev.} \textbf{\bibinfo{volume}{138}}, \bibinfo{pages}{A1421}
  (\bibinfo{year}{1965}).

\bibitem[{not0()}]{fn0}\bibinfo{note}{Early references on superconductivity
in narrow channels through films of OAPP are:
N. S. Enikolopyan, L. N. Grigorov, and S. G. Smirnova,
Pis'ma Zh. Eksp. Teor. Fiz. {\bf 49}, 326 (1989) [JETP Lett.
{\bf 49}, 371 (1989)];  
V. M. Arkhangorodski\u{i}, A. N. Ionov, V. M. Tuchkevich, 
and I. S. Shlimak, Pis'ma Zh. Eksp. Teor. Fiz. {\bf 51}, 56 (1990)
[JETP Lett. {\bf 51}, 67 (1990)];
O. V. Demicheva, D. N. Rogachev, S. G. Smirnova, E. I. Shklyarova,
M. Yu. Yablokov, V. M. Andreev, and L. N. Grigorov, Pis'ma Zh. Eksp. Teor.
Fiz. {\bf 51}, 228 (1990) [JETP Lett. {\bf 51}, 258 (1990)]. 
See Ref.~\onlinecite{Ea94_2} and D. M. Eagles, J. Supercond. {\bf 15}, 243 (2002)
for later references.}

\bibitem[{not()}]{fn}\bibinfo{note}{The authors of Ref.~\onlinecite{Zh05}
 discuss results on superconductivity at 
  room temperature in oxidised atactic polypropylene in terms of a model
of two-site bipolarons on adjoining one-dimensional chains.  However, 
we disagree with several results of these authors in
this and previous papers of theirs.}

\bibitem[{\citenamefont{Grigorov}(1990)}]{Gr90}
\bibinfo{author}{\bibfnamefont{L.~N.} \bibnamefont{Grigorov}},
  \bibinfo{journal}{Makromol. Chem., Macromol. Symp.}
  \textbf{\bibinfo{volume}{37}}, \bibinfo{pages}{159} (\bibinfo{year}{1990}).

\bibitem[{\citenamefont{Grigorov et~al.}(1990)\citenamefont{Grigorov, Andreev,
  and Smirnova}}]{GrAnSm90}
\bibinfo{author}{\bibfnamefont{L.~N.} \bibnamefont{Grigorov}},
  \bibinfo{author}{\bibfnamefont{V.~M.} \bibnamefont{Andreev}},
  \bibnamefont{and} \bibinfo{author}{\bibfnamefont{S.~G.}
  \bibnamefont{Smirnova}}, \bibinfo{journal}{Makromol. Chem., Macromol. Symp.}
  \textbf{\bibinfo{volume}{37}}, \bibinfo{pages}{177} (\bibinfo{year}{1990}).

\bibitem[{\citenamefont{Grigorov}(1991)}]{Gr91}
\bibinfo{author}{\bibfnamefont{L.~N.} \bibnamefont{Grigorov}},
  \bibinfo{journal}{Pis'ma Zh. Tekh. Fiz.} \textbf{\bibinfo{volume}{17}}
  \bibinfo{pages}{(5), 45} (\bibinfo{year}{1991}), \bibinfo{note}{[Sov. Tech. Phys.
  Lett. {\bf 17}, 368 (1991)]}.

\bibitem[{\citenamefont{McDonald and Ward}(1961)}]{McWa61}
\bibinfo{author}{\bibfnamefont{M.~P.} \bibnamefont{McDonald}} \bibnamefont{and}
  \bibinfo{author}{\bibfnamefont{I.~M.} \bibnamefont{Ward}},
  \bibinfo{journal}{Polymer} \textbf{\bibinfo{volume}{2}}, \bibinfo{pages}{241}
  (\bibinfo{year}{1961}).

\bibitem[{num(1986)}]{numrec}
\bibinfo{author}{\bibfnamefont{W.~H.} \bibnamefont{Press}},
\bibinfo{author}{\bibfnamefont{B.~P.} \bibnamefont{Flannery}},
\bibinfo{author}{\bibfnamefont{S.~A.} \bibnamefont{Teukolsky}},
\bibnamefont{and}
\bibinfo{author}{\bibfnamefont{W.~T.} \bibnamefont{Vetterling}},
\emph{\bibinfo{title}{Numerical Recipes}} (\bibinfo{publisher}{Cambridge
  University Press}, \bibinfo{year}{1986}).

\bibitem[{\citenamefont{Eagles}(2005)}]{Ea05}
\bibinfo{author}{\bibfnamefont{D.~M.} \bibnamefont{Eagles}},
  \bibinfo{journal}{Phil. Mag.} \textbf{\bibinfo{volume}{85}},
  \bibinfo{pages}{1931} (\bibinfo{year}{2005}).

\bibitem[{\citenamefont{Adhikari et~al.}(2000)\citenamefont{Adhikari, Casas,
  Puente, Rigo, Fortes, Sol\'{i}s, {De Llano}, Valladares, and
  Rojo}}]{AdCaPuRiFoSoDLVaRo00}
\bibinfo{author}{\bibfnamefont{S.~K.} \bibnamefont{Adhikari}},
  \bibinfo{author}{\bibfnamefont{M.}~\bibnamefont{Casas}},
  \bibinfo{author}{\bibfnamefont{A.}~\bibnamefont{Puente}},
  \bibinfo{author}{\bibfnamefont{A.}~\bibnamefont{Rigo}},
  \bibinfo{author}{\bibfnamefont{M.}~\bibnamefont{Fortes}},
  \bibinfo{author}{\bibfnamefont{M.~A.} \bibnamefont{Sol\'{i}s}},
  \bibinfo{author}{\bibfnamefont{M.}~\bibnamefont{{de Llano}}},
  \bibinfo{author}{\bibfnamefont{A.~A.} \bibnamefont{Valladares}},
  \bibnamefont{and} \bibinfo{author}{\bibfnamefont{O.}~\bibnamefont{Rojo}},
  \bibinfo{journal}{Phys. Rev. B} \textbf{\bibinfo{volume}{62}},
  \bibinfo{pages}{8671} (\bibinfo{year}{2000});
  \bibinfo{note} {Physica C {\bf 351}, 341 (2001)}.
  

\bibitem[{die()}]{dielectric}
\bibinfo{note}{Obtained from the square of the refractive index $n=1.49$ given
  by J. Brandrup and E. H. Immergut in {\em Polymer Handbook} (Wiley, New York,
  1989, Vol. V, p 27).}

\bibitem[{\citenamefont{Chen et~al.}(1998)\citenamefont{Chen, Ren, and
  Jiao}}]{ChReJi98}
\bibinfo{author}{\bibfnamefont{Q.~H.} \bibnamefont{Chen}},
  \bibinfo{author}{\bibfnamefont{Y.~H.}~\bibnamefont{Ren}}, \bibnamefont{and}
  \bibinfo{author}{\bibfnamefont{Zh.~K.}~\bibnamefont{Jiao}},
  \bibinfo{journal}{Eur. Phys. J. B}  
  \textbf{\bibinfo{volume}{3}},
  \bibinfo{pages}{307} (\bibinfo{year}{1998}).

\bibitem[{\citenamefont{Schultz}(1959)}]{Sc59}
\bibinfo{author}{\bibfnamefont{T.~D.} \bibnamefont{Schultz}},
  \bibinfo{journal}{Phys. Rev.} \textbf{\bibinfo{volume}{116}},
  \bibinfo{pages}{526} (\bibinfo{year}{1959}).

\bibitem[{\citenamefont{Romero et~al.}(1999)\citenamefont{Romero, Brown, and
  Lindenberg}}]{RoBrLi99I}
\bibinfo{author}{\bibfnamefont{A.~H.} \bibnamefont{Romero}},
  \bibinfo{author}{\bibfnamefont{D.~W.} \bibnamefont{Brown}}, \bibnamefont{and}
  \bibinfo{author}{\bibfnamefont{K.}~\bibnamefont{Lindenberg}},
  \bibinfo{journal}{Phys. Rev. B} \textbf{\bibinfo{volume}{59}},
  \bibinfo{pages}{13728} (\bibinfo{year}{1999}).

\bibitem[{n3()}]{n3}\bibinfo{note}{We have been informed that an
equivalent formula was obtained earlier by A.A. Gogolin, but we do not
have the reference}.

\bibitem[{\citenamefont{Toyozawa}(1961)}]{To61}
\bibinfo{author}{\bibfnamefont{Y.} \bibnamefont{Toyozawa}},
  \bibinfo{journal}{Prog. Theor. Phys. (Kyoto)}
  \textbf{\bibinfo{volume}{26}}, \bibinfo{pages}{29} (\bibinfo{year}{1961}).

\bibitem[{\citenamefont{Eagles}(1969)}]{Ea69}
\bibinfo{author}{\bibfnamefont{D.~M.} \bibnamefont{Eagles}},
  \bibinfo{journal}{Phys. Rev.}
  \textbf{\bibinfo{volume}{181}}, \bibinfo{pages}{1278} (\bibinfo{year}{1969}).

\bibitem[{n2()}]{n2}\bibinfo{note}
{Some errors in Ref.\onlinecite{Ea69} are noted in Sec.4 of D.M. Eagles and P. Lalousis,
J. Phys. C {\bf 17}, 655 (1984).}

\bibitem[{\citenamefont{Emin}(1973)}]{Em73}
\bibinfo{author}{\bibfnamefont{D.} \bibnamefont{Emin}},
  \bibinfo{journal}{Adv. Phys.}
  \textbf{\bibinfo{volume}{22}}, \bibinfo{pages}{57} (\bibinfo{year}{1973}).

\bibitem[{\citenamefont{L\'epine}(1992)}]{Le92}
\bibinfo{author}{\bibfnamefont{Y.} \bibnamefont{L\'epine}} \bibnamefont{and}
\bibinfo{author}{\bibfnamefont{Y.} \bibnamefont{Frongillo}},
  \bibinfo{journal}{Phys. Rev.}
  \textbf{\bibinfo{volume}{46}}, \bibinfo{pages}{14510} (\bibinfo{year}{1992}).

\bibitem[{\citenamefont{Iadonisi}(1998)}]{Ia98}
\bibinfo{author}{\bibfnamefont{G.} \bibnamefont{Iadonisi}},
\bibinfo{author}{\bibfnamefont{V.} \bibnamefont{Cataudella}},
\bibinfo{author}{\bibfnamefont{G.} \bibnamefont{de Filippis}}, 
\bibnamefont{and}
\bibinfo{author}{\bibfnamefont{D.} \bibnamefont{Ninno}},
  \bibinfo{journal}{Europhys. Lett.}
  \textbf{\bibinfo{volume}{41}}, \bibinfo{pages}{309} (\bibinfo{year}{1998}).

\bibitem[{\citenamefont{Eagles}(1999)}]{Ea99}
\bibinfo{author}{\bibfnamefont{D.~M.} \bibnamefont{Eagles}},
  \bibinfo{journal}{Revista Mexicana de Fisica, Suplemento 1}
  \textbf{\bibinfo{volume}{45}}, \bibinfo{pages}{118} (\bibinfo{year}{1999}).

\bibitem[{\citenamefont{Eagles}(1998)}]{Ea98}
\bibinfo{author}{\bibfnamefont{D.~M.} \bibnamefont{Eagles}},
  \bibinfo{journal}{Physica C} \textbf{\bibinfo{volume}{301}},
  \bibinfo{pages}{165} (\bibinfo{year}{1998}).

\bibitem[{\citenamefont{Pieri and Strinati}(2005)}]{PiSt05}
\bibinfo{author}{\bibfnamefont{P.}~\bibnamefont{Pieri}}, 
\bibinfo{author}{\bibfnamefont{L.}~\bibnamefont{Pisani}}, \bibnamefont{and}
  \bibinfo{author}{\bibfnamefont{G.}~\bibnamefont{Strinati}},
  \bibinfo{journal}{Phys. Rev. B} \textbf{\bibinfo{volume}{70}},
  \bibinfo{pages}{094508} (\bibinfo{year}{2004}).

\end{thebibliography}

\end{document}